\documentclass[reprint,superscriptaddress,showpacs,aps,prb]{revtex4-1}
\usepackage[pdftex]{graphicx}
\usepackage{amsmath}
\usepackage{amssymb}
\usepackage{placeins}
\usepackage{color}
\usepackage{xfrac}
\usepackage[letterpaper, total={7.28in, 9.55in}]{geometry}
\usepackage{pdfpages} 
\makeatletter 
\AtBeginDocument{\let\LS@rot\@undefined}
\makeatother
\usepackage{pgffor}

\usepackage{hyperref}
\usepackage{xcolor}
\hypersetup{
    colorlinks,
    linkcolor={red!50!black},
    citecolor={blue!50!black},
    urlcolor={blue!80!black}
}
\newcommand{\etal}{\textit{et al.}}
\newcommand{\muSR}{$\mu$SR}
\newcommand{\eis}{Eu$_5$In$_2$Sb$_6$}
\newcommand{\qG}{$\mathbf{q}_\mathit{\Gamma}$}

\newcommand{\G}{$\mathit{\Gamma}$}
\newcommand{\Z}{$Z$}

\newcommand{\Gt}{$\mathit{\Gamma}_3^+$} 

\newcommand{\Zt}{$Z_3^+$} 

\newcommand{\TZ}{$T_Z$}
\newcommand{\TN}{$T_\mathrm{N}$}

\begin{document}

\graphicspath{{.}{figures/}}

\title{Unusual magnetism of the axion-insulator candidate Eu$_5$In$_2$Sb$_6$}

\author{M. C. Rahn}
\email[]{marein.rahn@tu-dresden.de}
\affiliation{Institute for Solid State and Materials Physics, Technical University of Dresden, 01062 Dresden, Germany}
\affiliation{Los Alamos National Laboratory, Los Alamos, New Mexico 87545, USA}
\author{M. N. Wilson}
\affiliation{Department of Physics, Durham University, South Road, Durham, DH1 3LE, United Kingdom}
\affiliation{Memorial University, Department of Physics and Physical Oceanography, St. John's, NL, A1B\,3X7, Canada}
\author{T. J. Hicken}
\affiliation{Department of Physics, Durham University, South Road, Durham, DH1\,3LE, United Kingdom}
\affiliation{Laboratory for Muon-Spin Spectroscopy, Paul Scherrer Institute, CH-5232 Villigen, Switzerland}
\author{F.L. Pratt}
\affiliation{ISIS Facility, STFC, Rutherford Appleton Laboratory, Chilton, Didcot, Oxfordshire, OX11\,0QX, United Kingdom}
\author{C. Wang}
\affiliation{Laboratory for Muon-Spin Spectroscopy, Paul Scherrer Institute, CH-5232 Villigen, Switzerland}
\author{F. Orlandi}
\affiliation{ISIS Facility, STFC, Rutherford Appleton Laboratory, Chilton, Didcot, Oxfordshire, OX11\,0QX, United Kingdom}
\author{D. D. Khalyavin}
\affiliation{ISIS Facility, STFC, Rutherford Appleton Laboratory, Chilton, Didcot, Oxfordshire, OX11\,0QX, United Kingdom}
\author{P. Manuel}
\affiliation{ISIS Facility, STFC, Rutherford Appleton Laboratory, Chilton, Didcot, Oxfordshire, OX11\,0QX, United Kingdom}
\author{L. S. I. Veiga}
\affiliation{London Centre for Nanotechnology and Department of Physics and Astronomy, University College London, London WC1E\,6BT, United Kingdom}
\affiliation{Diamond Light Source Ltd., Didcot OX11\,0DE, United Kingdom }
\author{A. Bombardi}
\affiliation{Diamond Light Source Ltd., Didcot OX11\,0DE, United Kingdom }
\author{S. Francoual}
\affiliation{Deutsches Elektronen-Synchrotron (DESY), Hamburg, Germany}
\author{P. Bereciartua}
\affiliation{Deutsches Elektronen-Synchrotron (DESY), Hamburg, Germany}
\author{A. S. Sukhanov}
\affiliation{Institute for Solid State and Materials Physics, Technical University of Dresden, 01062 Dresden, Germany}
\author{J. D. Thompson}
\affiliation{Los Alamos National Laboratory, Los Alamos, New Mexico 87545, USA}
\author{S. M. Thomas}
\affiliation{Los Alamos National Laboratory, Los Alamos, New Mexico 87545, USA}
\author{P. F. S. Rosa}
\affiliation{Los Alamos National Laboratory, Los Alamos, New Mexico 87545, USA}
\author{T. Lancaster}
\affiliation{Department of Physics, Durham University, South Road, Durham, DH1\,3LE, United Kingdom}
\author{F. Ronning}
\affiliation{Los Alamos National Laboratory, Los Alamos, New Mexico 87545, USA}
\author{M. Janoschek}
\affiliation{Los Alamos National Laboratory, Los Alamos, New Mexico 87545, USA}
\affiliation{Laboratory for Neutron and Muon Instrumentation, Paul Scherrer Institute, CH-5232 Villigen, Switzerland}
\affiliation{Physik-Institut, Universität Zürich, CH-8057 Zürich, Switzerland}
\email[]{marc.janoschek@psi.ch}

\date{\today}

\begin{abstract}

\eis~is a member of a family of orthorhombic nonsymmorphic rare-earth intermetallics that combines large localized magnetic moments and itinerant exchange with a low carrier density and perpendicular glide planes. This may result in special topological crystalline (wallpaper fermion) or axion insulating phases. Recent studies of \eis~single crystals have revealed colossal negative magnetoresistance and multiple magnetic phase transitions. Here, we clarify this ordering process using neutron scattering, resonant elastic X-ray scattering, muon spin-rotation, and magnetometry. The nonsymmorphic and multisite character of \eis~results in coplanar noncollinear magnetic structure with an Ising-like net magnetization along the $a$ axis. A reordering transition, attributable to competing ferro- and antiferromagnetic couplings, manifests as the onset of a second commensurate Fourier component. In the absence of spatially resolved probes, the experimental evidence for this low-temperature state can be interpreted either as an unusual double-$q$ structure or in a phase separation scenario. The net magnetization produces variable anisotropic hysteretic effects which also couple to charge transport. The implied potential for functional domain physics and topological transport suggests that this structural family may be a promising platform to implement concepts of topological antiferromagnetic spintronics.
\end{abstract}

\maketitle

\section{Introduction}

The guided design and control of topologically protected conduction states on surfaces and interfaces of bulk insulators and semimetals is one of the most sought-after results of quantum materials physics. The concept of topological quantum chemistry~\cite{Bradlyn2017} has enabled high-throughput calculations and predictions of topological band properties in the absence of electronic correlations~\cite{Zhang2019,Vergniory2019}. While progress on intrinsically magnetic topological matter has been much slower, the analysis of all magnetic space groups and tabulated magnetic structures indicates that there is indeed an abundance of candidate strongly correlated materials with potential for functional band-topology~\cite{Watanabe2018,Xu2020}. However, as magnetic symmetry can currently not be predicted from first principles, the identification of suitable material platforms remains an outstanding challenge.

Following the proposal of fourfold degenerate Dirac surface states with hourglass connectivity in the non-symmorphic orthorhombic Zintl phase Ba$_5$In$_2$Sb$_6$~\cite{Wieder2018}, there has been an increased interest in magnetic members of this 5-2-6 structural family. The narrow band-gap semiconductor \eis~had been previously studied for its outstanding thermoelectric properties at high temperature~\cite{Chanakian2015,Lv2017}. Following the synthesis of single-crystalline samples, the material revealed highly unusual low-temperature properties, including colossal magnetoresistance, which has been attributed to the localization of charge carriers at magnetic polarons~\cite{Rosa2020,Souza2022,Ghosh2022}. In-depth density functional studies of Eu-based 5-2-6 compounds have since identified the potential for axion insulating phases and predicted a high sensitivity to chemical and pressure/strain tuning~\cite{Varnava2022}.  However, both the electronic and magnetic structures of the 5-2-6 compounds have so far eluded experimental determination.

\begin{figure}
\includegraphics[width=0.96\columnwidth,trim= 0pt 0pt 0pt 0pt, clip]{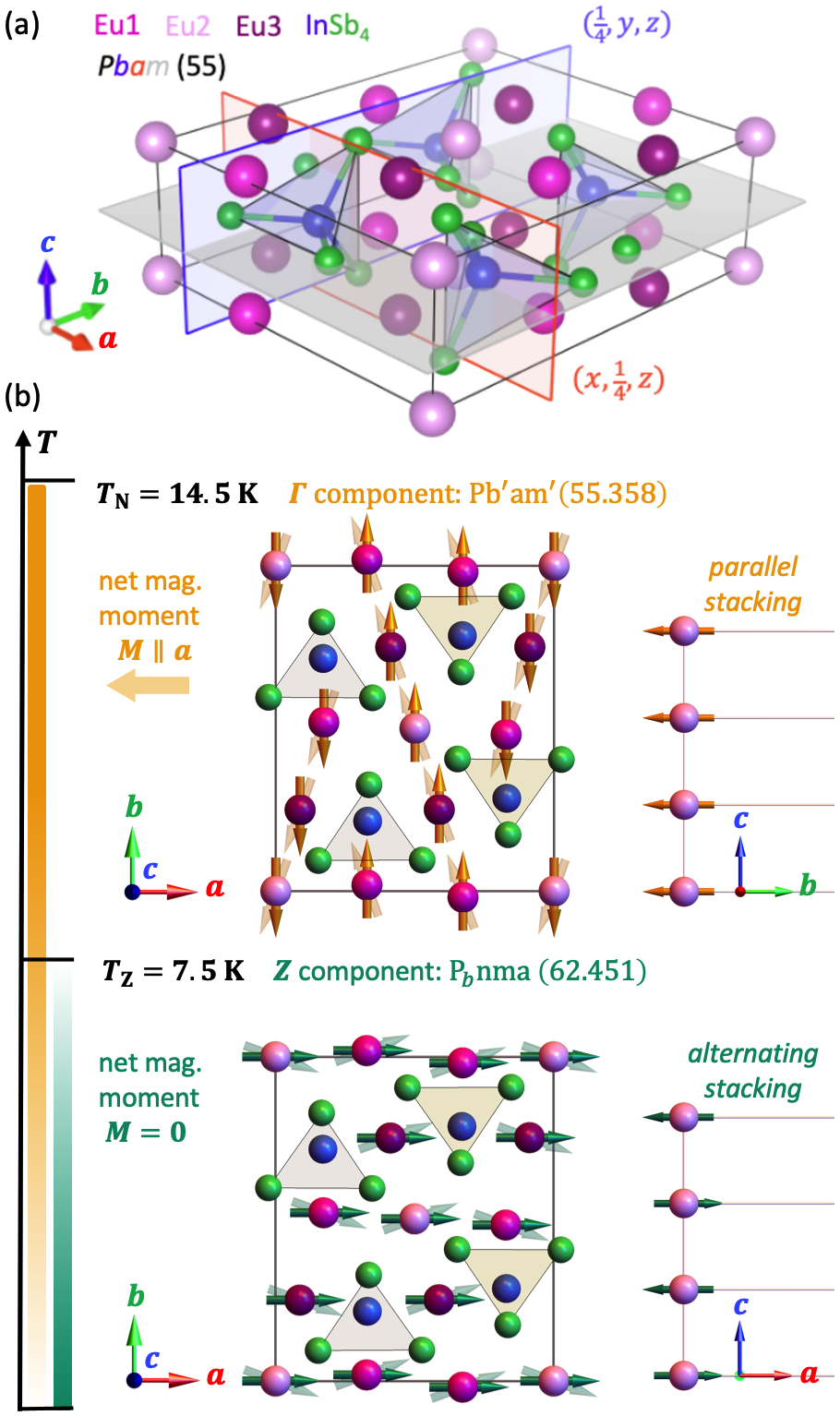}
\caption{\label{Fig1} (a) The orthorhombic unit cell and $Pbam$ symmetries of \eis. The perpendicular glide planes are highlighted in blue and red. (b) Solid arrows show the spin arrangement inferred for the \G~magnetic component, which forms at $T_\mathrm{N}=14.5$\,K. The structure is collinear within the uncertainty of the neutron diffraction data. Shaded arrows illustrate tilts away from this collinear configuration, which creates a net magnetization. (c) Corresponding views of the \Z~magnetic component, which originates below $T_Z=7.5$\,K. The X-ray and neutron scattering data do not allow the distinction between a double-$q$ state and magnetic phase separation. Both \G~and \Z~components are described by the same magnetic basis vectors, which generally add to a per-layer magnetization along the $a$ axis. For the \G~component, this results in a spontaneous macroscopic magnetization, while in the \Z~component it is compensated by alternating stacking.}
\end{figure}

Here, we report on the complex magnetic ordering process of \eis. Using muon spin-rotation ($\mu$SR), neutron powder diffraction (NPD) and resonant elastic x-ray scattering (REXS), we are able to shed light on an unusual two-step ordering process due to the nonsymmorphic and multi-site character of \eis. Unexpectedly, our findings imply the existence of net-magnetized domains, and angle-dependent electrical resistivity measurements indicate coupling of magnetism and charge transport.

Fig.~\ref{Fig1}(a) illustrates the orthorhombic $Pbam$ unit cell of \eis. The structure can be understood as layers of coplanar Eu ions (ten per unit cell) that are interpenetrated by chains of corner-sharing InSb$_4$ tetrahedra along the $c$ axis. The figure also illustrates the positions of the $b$ glide (blue), the $a$ glide (red), and the mirror plane (gray). The perpendicular glide planes endow the Eu-layers with the symmetry of the wallpaper group $pgg$, encountered in daily life as the herringbone pattern~\cite{SM}. This combination of non-symmorphic symmetries is the prerequisite for the novel topological-crystalline phases predicted in this structural family~\cite{Wieder2018}. In Eu$_5$$M_2X_6$ compounds~\cite{Varnava2022}, this potential for non-trivial topology combines with the large magnetic moment of divalent europium ($S=7/2$, $L=0$, $\mu\approx8\mu_\mathrm{B}$).

We find that below the Néel temperature $T_\mathrm{N}=14.5$\,K, the Eu$^{2+}$ spins initially form a complex coplanar, non-collinear arrangement (``\G'') that carries a net magnetic moment along the $a$ axis, as illustrated in Fig.~\ref{Fig1}(b). Below $T_Z=7.5$\,K, this phase is either gradually displaced by a growing volume fraction of another magnetic phase (``\Z''), or it forms a double-$q$ state with perpendicular \G~and \Z~components. Crucially, the per-layer spin arrangements of the \G~and \Z~components obey the same magnetic symmetries -- they differ only in the (parallel vs. alternating) stacking of this motif along the $c$axis. While all three magnetic states (\G, \Z, double-$q$) conserve inversion symmetry $\mathcal{I}$, effective time reversal symmetry $\{\mathcal{T}|00\frac12\}$ is only conserved in the \Z~state, which would make these regions candidate axion insulators~\cite{Sekine2021,Bernevig2022}. The phase separation scenario could therefore provide a platform to explore interfaces of topologically distinct insulating regimes.

\section{Results and interpretation}

\begin{figure}
\includegraphics[width=1.0\columnwidth,trim= 0pt 0pt 0pt 0pt, clip]{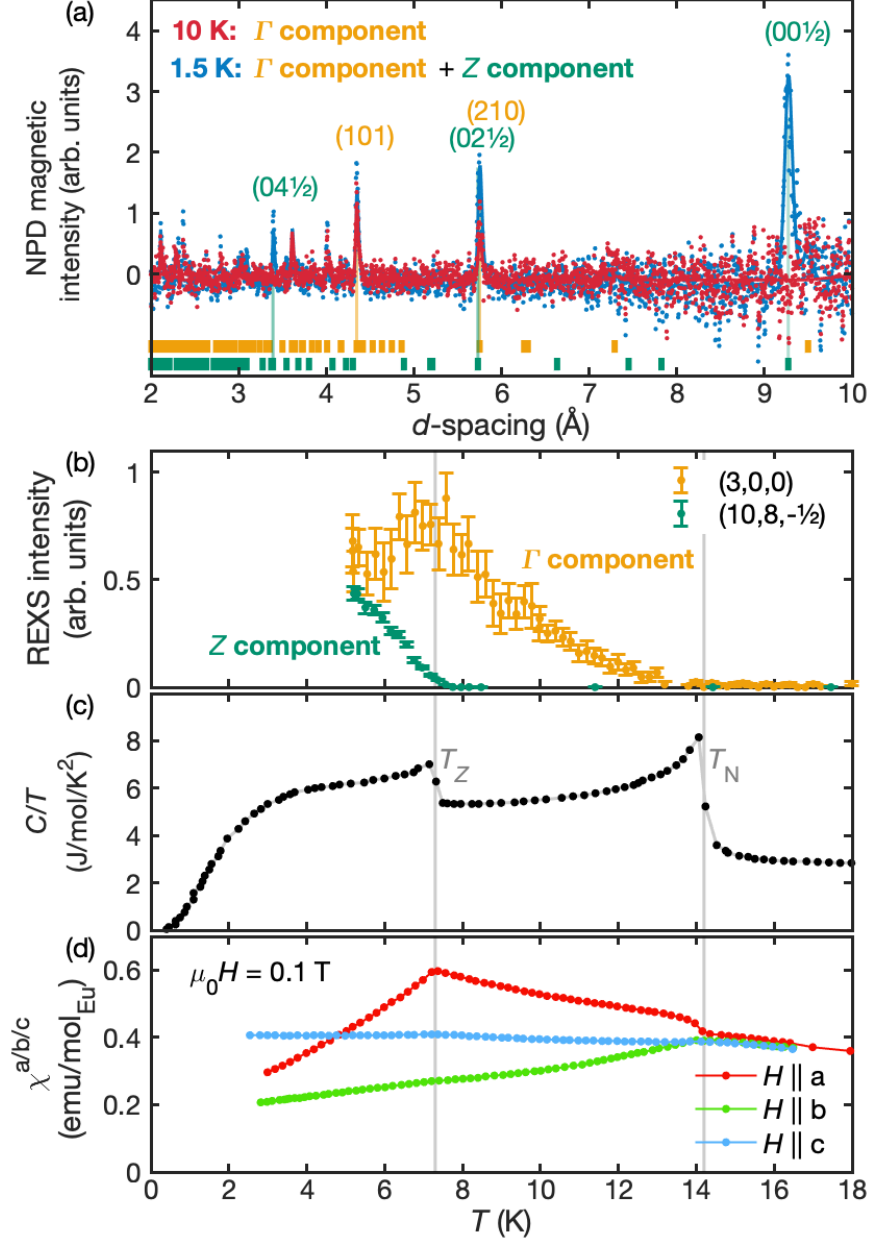}
\caption{\label{Fig2} (a) Magnetic neutron powder diffraction at 1.5\,K and 10\,K. The nuclear intensity (at 25\,K) has been subtracted from the data~[\onlinecite{SM}]. (b) Temperature dependence of REXS intensities associated with the propagation vectors of the Ising weakly-ferromagnetic \G~order and the compensated antiferromagnetic \Z~component. Below $T_Z=7.5$\,K, $\mathit{\Gamma}$ and $Z$ either coexist by phase separation or as a double-$q$ structure. (c) The consecutive second-order phase transitions at 14.5\,K and 7.5\,K observed in heat capacity ($C/T$). (d) The anisotropic magnetic susceptibility $\chi^{a/b/c}(T)$ at 100\,mT applied along the principal axes. Note the small spontaneous magnetization of $\chi^a$ in the \G~state, and the change in easy axis below $T_Z$.} 
\end{figure}

Previous measurements of the heat capacity and magnetic susceptibility~\cite{Rosa2020}, reproduced in Fig.~\ref{Fig2}(c,d), revealed two continuous magnetic transitions, which correspond to the Néel temperature $T_\mathrm{N}=14.5\,$K, and the onset of the \Z~component at $T_Z=7.5\,$K. The magnetic neutron powder diffraction patterns in Fig.~\ref{Fig2}(a) illustrate that magnetic Bragg peaks in the intermediate phase ($T_Z<10\,\mathrm{K}<T_\mathrm{N}$) can be indexed by the propagation vector $\mathbf{q}_\mathit{\Gamma}=(0,0,0)$. The low-temperature dataset ($1.5\,\mathrm{K}<T_Z$) features both  $\mathbf{q}_\mathit{\Gamma}$ and $\mathbf{q}_Z=(0,0,\sfrac12)$ peaks~\cite{SM}.

Using resonant elastic x-ray scattering (REXS) at the Eu $L$ edges, we characterized the $\mathbf{q}_Z$ and $\mathbf{q}_\mathit{\Gamma}$ magnetic order in more detail. Fig.~\ref{Fig2}(b) shows the thermal evolution of Bragg intensities associated with the two Fourier components. These intensities are not necessarily a direct measure of the ordered magnetic moment because the magnetic order has degrees of freedom that affect the structure factor and may evolve with temperature. Nevertheless, the rise of $\mathbf{q}_Z$ intensity below $T_Z$ appears to be proportional to the reduction of the $\mathbf{q}_\mathit{\Gamma}$ peak, consistent with the displacement of the \G~by the \Z-type Fourier component. While for a double-$q$ model the \G/\Z~ratio may be different at each Wyckoff site, the refinement of the 1.5\,K neutron powder diffraction data in the phase separation scenario results in phase volumes of 34\% (\Z) and 66\% (\G)~\cite{SM}.

Azimuthal scans and full linear polarisation analysis of the REXS signal, shown in the Supplemental Material~\cite{SM}, reveal the direction of magnetic structure factor vectors of individual Bragg peaks. Combined with representational analysis, this information determines the irreducible representations (irreps) associated with the two magnetic phases~\cite{SM}. The \G~and \Z~magnetic structures are described by the irreps \Gt~($Pb'am'$) and \Zt ($P_{b}nma$), and the superposition of both would correspond to magnetic space group $Pb'am'$. Crucially, these two irreps impose the same phase relations (basis vectors) on the relative alignment of the coplanar magnetic moments -- the two magnetic components differ only in the parallel (\G) or alternating (\Z) stacking of this motif along the $c$ axis. Given that magnetometry reveals a localized $S=7/2$ ($\approx8\,\mu_\mathrm{B}$) magnetic moment at each site~\cite{Rosa2020}, a double-$q$ structure can only form under the special condition that the Fourier components associated with either irrep must be perpendicular to one another (else, the magnitude of the moments would alternate between layers). As indicated by solid arrows in Fig.~\ref{Fig1}(b,c), we find that the magnetic components associated with the \G~and \Z~magnetic structures are close to collinear with the $b$ and $a$ axes, respectively. The shaded arrows illustrate tilts away from the collinear structures that are allowed by symmetry and evidenced by magnetometry. The magnitude of the tilts is either smaller than the uncertainty of the data (neutron scattering) or $1^\circ\sim10^\circ$ (REXS, \Z~component).

The per-layer arrangement of the ten magnetic moments has several important implications. The Eu ions occupy one 2-fold and two 4-fold  Wyckoff sites, as indicated by different shades in Fig.~\ref{Fig1} (see [\onlinecite{SM}] for more detailed illustrations). The overall structure is defined by three in-plane spin orientations (one per site), which may vary continuously with temperature.  Crucially, the components of the magnetic moments (anti-)parallel to the $b$ axis cancel necessarily, but all $a$ axis components per Wyckoff orbit are parallel. Given the 2/4/4 multiplicity, they can, therefore, in general, not be compensated between the three sites. This makes each layer a weak (canted) Ising ferromagnet with a variable net magnetization along the $a$ axis. Notably, this holds true independently of whether the moments are mostly aligned with the $a$ or $b$ axis, as shown in Fig.~\ref{Fig1}(b). For instance, in the \G~component, the net magnetization of layers is small because the moments are, within the accuracy of our measurements, aligned with the $b$ axis. However, small tilts away from this collinear configuration (indicated by shaded arrows) produce a small net moment along $a$, which adds up to a macroscopic magnetization. Conversely, the \Z~magnetic component carries a large per-layer magnetization along $a$ (the spins are close to collinear along $a$), but here this is compensated by the alternating stacking.

Knowledge of this magnetic order explains several unusual observations of bulk and local magnetic probes. For example, in Fig.~\ref{Fig2}(d), the $H\parallel a$ magnetic susceptibility $\chi^a(T)$ shows a ferromagnet-like increase below $T_{\mathrm{N}}$. This is not compatible with conventional antiferromagnetism but can now be explained by the weak $M\parallel a$ ferromagnetism. Below $T_Z$, the bulk susceptibility gradually takes on the character of the $Z$ magnetic structure shown in Fig.~\ref{Fig1}(b): As the weakly ferromagnetic (\G) component recedes, $\chi^a(T)$ decreases rapidly. Eventually (below $\approx5$\,K), the magnetic susceptibility is largest for $H\parallel c$, as expected of a coplanar antiferromagnet. Given that the moments of the \Z~component align largely with the $a$ axis, see Fig.~\ref{Fig1}(b), $\chi^a$ should be minimal once the \Z~component dominates. Accordingly, the gradient of $\chi^a(T)$ below $T_Z$ is much steeper than that of $\chi^b(T)$. 

\begin{figure}[h!]
\includegraphics[width=1.0\columnwidth,trim= 0pt 0pt 0pt 0pt, clip]{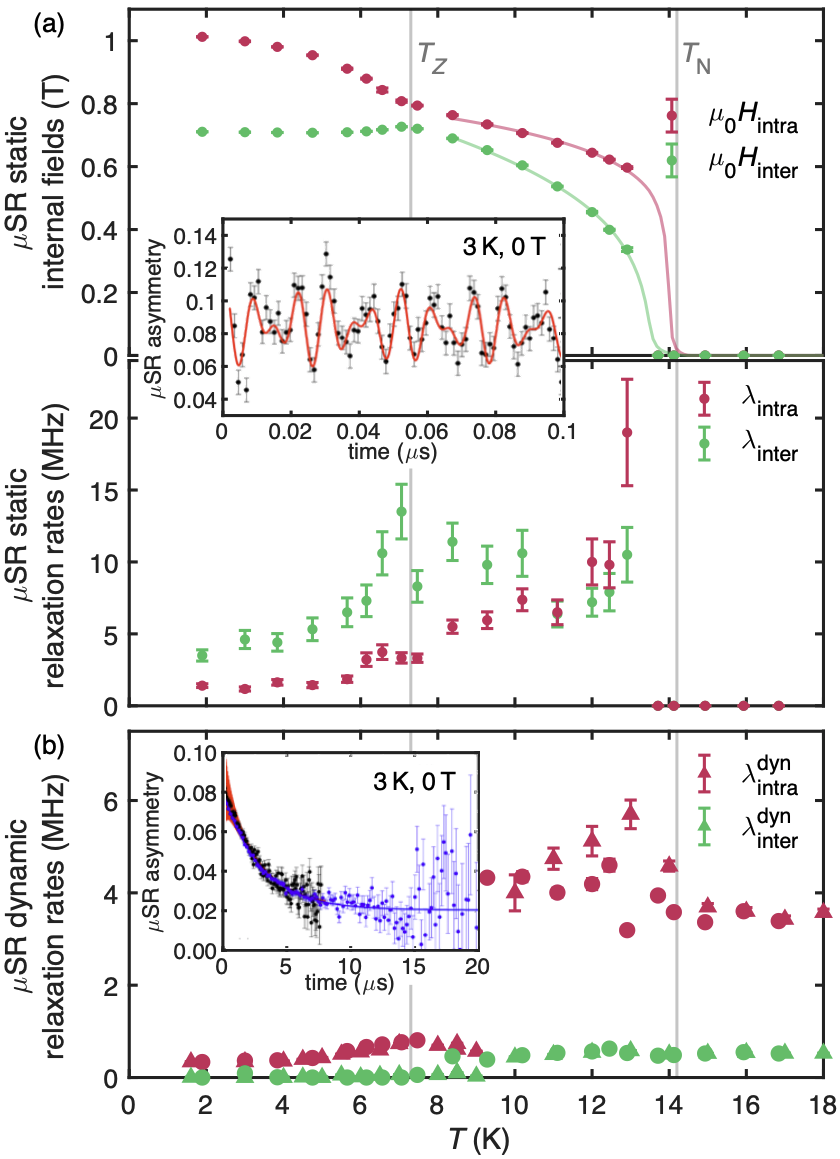}
\caption{\label{FigMuSR}  (a) Static internal fields (top) and relaxation rates (bottom) determined by zero-field continuous-source $\mu$SR (GPS instrument). Solid lines are a guide to the eye. We attribute different components of the $\mu$SR asymmetry to muon stopping sites that are more sensitive to fields within (intra) and between (inter) the layers. The inset shows an example of the $\mu$SR asymmetry measured at GPS (3\,K, typical fit indicated by a red line). (b) Dynamic relaxation rates determined from the decay of the zero-field $\mu$SR asymmetry up to 20\,$\mu$s. Circles and triangles mark values inferred from data measured at continuous (GPS) and pulsed source (EMU) muon instruments, respectively. The inset shows a typical fit in this time window, including both continuous source (black) and pulsed source data (blue).}
\end{figure}

To probe the magnetic ordering process from a local point of view, we turn to muon spin-rotation ($\mu$SR) experiments. In order to reveal both static and dynamic characteristics ($\mu$s timescale), we measured spectra at continuous (PSI/GPS) and pulsed (ISIS/EMU) muon sources. The observed $\mu$SR asymmetry can be interpreted by assuming two types of muon stopping sites. Each contributes an oscillating component associated with a characteristic relaxation rate $\lambda$ and a dynamical relaxation $\lambda^\mathrm{dyn}$ (due to fluctuating fields). One possible interpretation of the fit results is that one class of stopping sites is more sensitive to magnetic fields within the planes (intra), and the other is more sensitive to fields between the planes (inter). The variation of these characteristics throughout the ordering process, along with exemplary fits to the data, is shown in Fig.~\ref{FigMuSR}.

As seen in Fig.~\ref{FigMuSR}(a), the signal that we associate with static in-plane fields $H_\mathrm{intra}$ sets in abruptly at \TN, as often observed for materials with two-dimensional correlations. By contrast, the gradual onset of $H_\mathrm{inter}$ is more typical for three-dimensional magnetism. Interestingly, as seen in Fig.~\ref{FigMuSR}(b), neither dynamical relaxation $\lambda_\mathrm{inter}^\mathrm{dyn}$, nor $\lambda_\mathrm{intra}^\mathrm{dyn}$, which encode magnetic fluctuations, have strong anomalies at \TN. The transition could therefore be interpreted as a freezing-out of pre-existing short-range magnetic correlations.

At \TZ, the absence of discontinuities of the static fields in Fig.~\ref{FigMuSR}(a) is consistent with the picture that the local magnetic structure does not change abruptly. Since the onset of an alternating stacking (\Z) component does not affect the per-layer magnetic symmetry, the static fields within the layers ($H_\mathrm{intra}$, $\lambda_\mathrm{intra}$) are not strongly affected by this transition. By contrast, the onset of the \Z~component qualitatively changes the trend of $H_\mathrm{inter}$ and causes an anomaly in $\lambda_\mathrm{inter}$. Moreover, as shown in Fig.~\ref{FigMuSR}(b), both dynamic relaxation rates ($\lambda_\mathrm{inter}^\mathrm{dyn}$, $\lambda_\mathrm{intra}^\mathrm{dyn}$) collapse around \TZ. This implies that the magnetism overall becomes much more static, possibly because competition between magnetic correlations is relieved. Below \TZ, $H_\mathrm{inter}$ has a slight tendency to decrease upon cooling. This unusual behavior may reflect changes in the domain structure, as has been previously suggested in GaV$_4$S$_8$~\cite{Hicken2020,Franke2018}. Notably, the relative amplitudes of the components do not change below \TZ, as one might expect for varying volume fractions. Although this might be taken as evidence for the double-$q$ scenario, the nature of the \G~and \Z~magnetic orders (which share the same magnetic symmetry per layer) means that we cannot discount the phase separation picture. 

\begin{figure}
\includegraphics[width=1\columnwidth,trim= 0pt 0pt 0pt 0pt, clip]{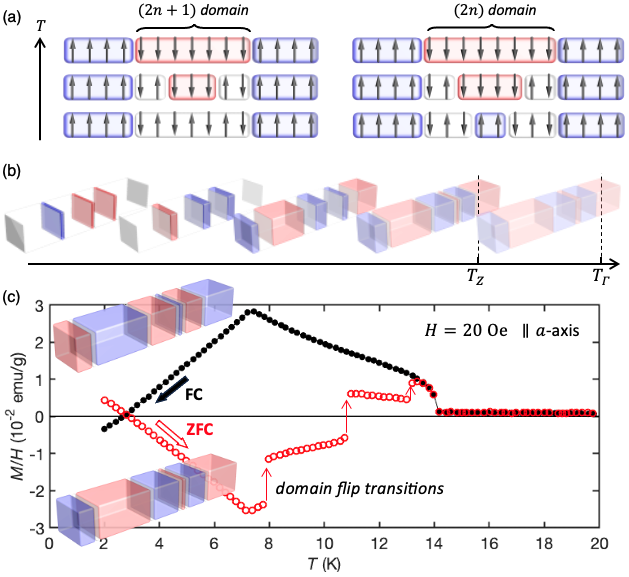}
\caption{\label{Fig3} (a) Illustration of the magnetic phase separation scenario below $T_Z$. Each layer of \eis~is represented by an arrow indicating the direction of the Ising (per-layer) net magnetic moment. Up (down) domains of the \qG~state are colored blue (red). Interfaces between \G~domains may provide natural nucleation points of the \Z~state (white). \G~domains with an odd number of layers can turn into a continuous $Z$ volume, while those with an even number of layers leave behind a soliton with a net magnetic moment. (b) Sketch of the thermal evolution of an \eis~slab, colored in analogy to (a). (c) Field-cooled and zero-field cooled (FC/ZFC) curves of the magnetic susceptibility measured with a weak (20\,Oe) bias field along the $a$ axis. Magnetization reversal and domain flip transitions [indicated by inset illustrations] provide a tentative explanation for these unusual characteristics.}
\end{figure}

Neither local or bulk magnetic probes nor magnetic scattering allow a clear distinction between a double-$q$ or phase separation scenario below \TZ. In either case, the net magnetization of the \G~state implies that Ising-like magnetic domains and their interplay on the mesoscale may play an important role in \eis, as illustrated in Fig.~\ref{Fig3}. For instance, as observed in other phase-separated magnets~\cite{Guo2022}, interfaces of ferromagnetic domains may provide natural nucleation points for the growth of antiferromagnetic (\Z) regimes. As illustrated in Fig.~\ref{Fig3}(a,b), \G~domains with an odd number of layers could be converted entirely into a \Z~region. By contrast, \G~domains with an even number of layers necessarily leave behind solitonic interfaces with net magnetization, potentially opposite to the original magnetization~\cite{Papanicolaou1995}. 

Such effects provide a compelling scenario to understand the unusual field and zero-field cooled (FC/ZFC) characteristics of the $H\parallel a$ magnetic susceptibility shown in Fig.~\ref{Fig3}(c). Upon warming, a ZFC sample in which, at low $T$, the opposite domain configuration happens to be encoded shows the same (FC) magnetic susceptibility curve of opposite sign until the thermal energy is sufficient to allow consecutive flips of \G~domains (or domains carrying a \G~component) into the field direction. The observed sign reversal of the magnetic susceptibility might be attributable to a specific \G-component domain structure dictated by demagnetization fields, which would also be compatible with the double-$q$ scenario. Details of this process should depend on the sample shape and pinning centres of antiferromagnetic domain walls, which could be verified by imaging techniques.

\begin{figure}
\includegraphics[width=1.0\columnwidth,trim= 0pt 0pt 0pt 0pt, clip]{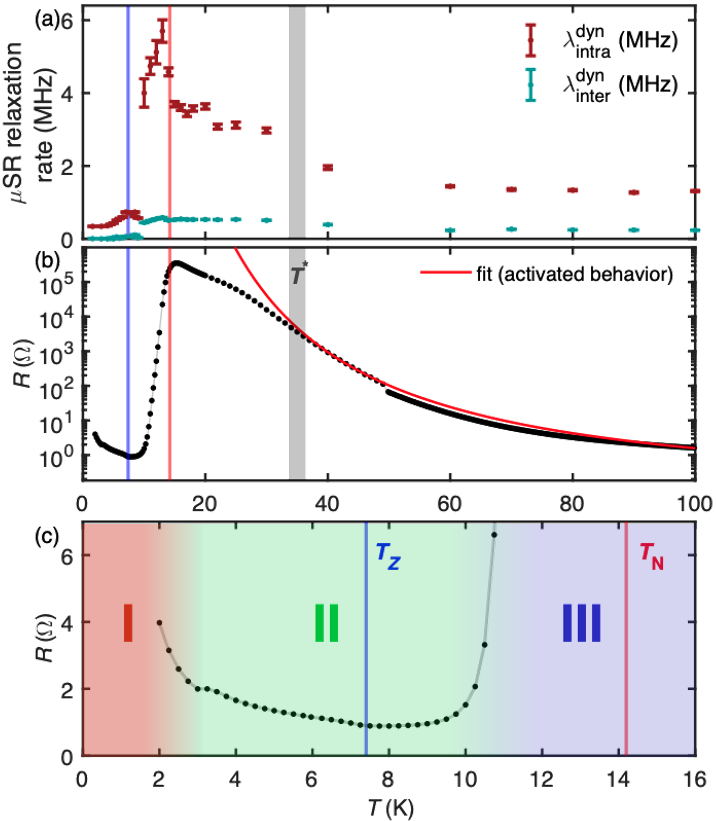}
\caption{\label{Fig4} (a) Pulsed-source \muSR~measurements of the dynamic relaxation rates $\lambda_\mathrm{intra}^\mathrm{dyn}$ and $\lambda_\mathrm{inter}^\mathrm{dyn}$ indicate an onset of short-ranged magnetic order at $T^*\approx 35$\,K. (b) On the same temperature scale, the resistance $R$ (current $I\parallel c$, log scale) begins to diverge from activated behavior. (c) Detailed view of the resistance within the magnetically ordered phase (linear scale). We identify three regimes of charge transport (I,II,III).}
\end{figure}

\begin{figure*}
\includegraphics[width=2.0\columnwidth,trim= 0pt 0pt 0pt 0pt, clip]{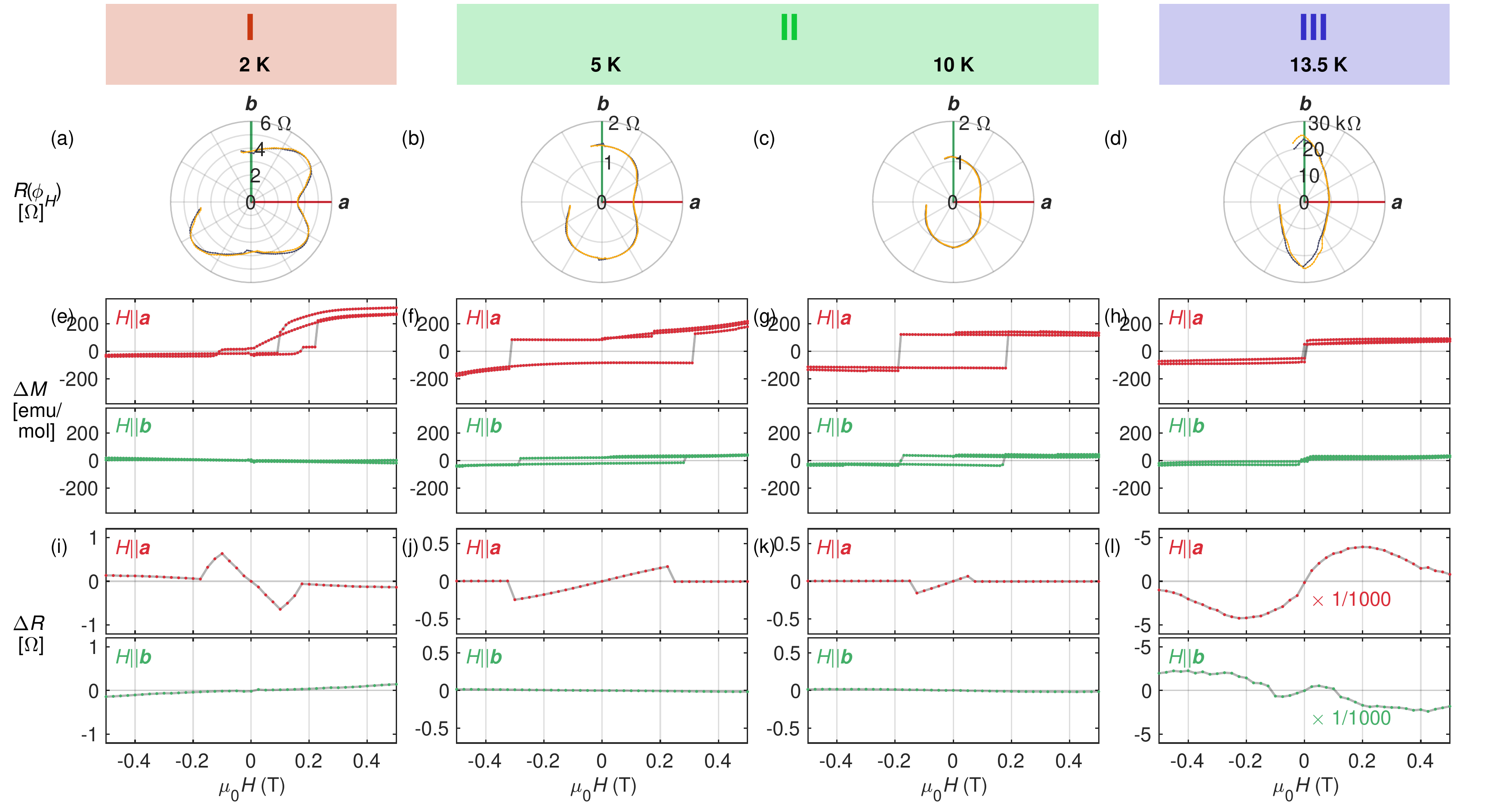}
\caption{\label{Fig5} Characteristic anisotropic hysteretic phenomena in the three regimes of charge transport (I, II, III). (a-d) Polar plots of the resistance $R$ at 1\,T (current $I\parallel\hat{c}$) for varying field directions $\phi_H$ in the $a$--$b$ plane. The three regimes differ not only in the magnitude of $R$ [cf. Fig.~\ref{Fig3}(c)], but also in their anisotropy. (e-h) Respective hysteretic components in the magnetization $\Delta M(H)$ for fields along $a$ and $b$. Linear slopes have been subtracted from each curve. (i-l) Corresponding hysteretic contributions $\Delta R(H)=R(H_\uparrow)-R(H_\downarrow)$ to the resistivity. Note both the quantitative and qualitative variation between the regimes. In regime II (3.0\,K$\sim$10\,K), $\Delta R(H)$ has a positive slope and is concomitant with $\Delta M$. It can therefore be associated with the (re-)orientation of magnetic domains of the \G~magnetic order.}
\end{figure*}

The magnetism of~\eis~appears to be coupled to charge transport over a wide temperature range. Fig.~\ref{Fig4}(a) illustrates that the \muSR~dynamic relaxation rates $\lambda_\mathrm{inter}^\mathrm{dyn}$ and $\lambda_\mathrm{intra}^\mathrm{dyn}$, associated with slow magnetic dynamics, increase significantly when cooling below $T^*\approx 35$\,K. This temperature scale has previously been associated with the onset of strong short-ranged antiferromagnetic correlations~\cite{Rosa2020}. As illustrated in Fig.~\ref{Fig4}(b), $T^*$ also coincides with the temperature below which the resistance diverges from activated behavior, $R\propto T\,\exp(42\,\mathrm{meV}/k_\mathrm{B}T)$~\cite{Rosa2020}. Taken together, this suggests that below $T^*$, the charge transport in \eis~starts being affected by magnetic dynamics.

Hysteretic characteristics in \eis~change throughout the magnetic ordering process, with a variable effect on charge transport. As illustrated in Fig.~\ref{Fig4}(c), we identify three temperature regimes of charge transport: The CMR regime above $T>10.5$\,K (III), the region 3.0\,K $<T<10.5$\,K where the resistance is reduced by more than five orders of magnitude (II), and a steep rise of resistance below $T<$ 3.5\,K (I). Figs.~\ref{Fig5}(a--d) show how the resistance (current $I\parallel\hat{c}$) varies under an azimuthal rotation in an applied field of 1\,T. The three transport regimes differ not only quantitatively (in the magnitude of $R$) but also qualitatively.

The corresponding Figs.~~\ref{Fig5}(e--l) show hysteretic effects for fields applied along the $a$ and $b$ axes, as observed in the magnetization (e--f) and resistance (i--l). To isolate the hysteretic components $\Delta M$ and $\Delta R$, a linear slope was subtracted from the $M(H)$ curves, and up- and down-ramps of $R(H)$ were subtracted. The observations are similar at 5\,K and 10\,K (i.e. below and above $T_Z$), see Fig.~\ref{Fig5}(f,g)/(j,k). Specifically, both $\Delta M$ and $\Delta R$ are negligible for $H\parallel\hat{b}$, but sizable and of coinciding coercive fields for $H\parallel\hat{a}$. This matches the expectations for the reversal of weakly ferromagnetic domains -- which are indeed expected below and above $T_Z$. Variations of the coercive field and remanent magnetization between 5\,K and 10\,K are expected, given the trade-off between a decreasing ordered magnetic moment and an increasing \G~Fourier component (or possibly \G~volume fraction). The linear $\Delta R(H)\propto \pm H$ at low temperatures distinguishes the hysteretic effects in regimes (I,II) against the much larger and non-monotonic $\Delta R(H)$ in the CMR regime (III). Interestingly, $\Delta R(H)$ changes sign between (I) and (II), but retains its association with the $a$ axis. 

\section{Discussion}
Our findings show that the magnetic ordering process in \eis~is shaped by its non-symmorphic orthorhombic structure and, specifically, by the fact that magnetic ions occupy multiple, non-equivalent Wyckoff sites. A search for rare earth intermetallics with these characteristics indeed yields several interesting materials with similarly complex magnetism. For instance, the $R_5$Ni$_2$In$_4$ structural family is also of $Pbam$ symmetry and also features a rare earth $R$ on two 4-fold and one 2-fold site. In contrast to the 5-2-6 family, since the $R$ ions are here trivalent, a range of rare earths can be substituted. For instance, a similar complex non-collinear magnetism with multiple phase transitions is found in for $R=$Tm~\cite{Szytula2014}, and similar characteristics, including weak ferromagnetism, can also be obtained above 100\,K by substituting $R=$Dy~\cite{Provino2012}.

Multiple magnetic (re-)ordering transitions at zero field are a common observation among these and related orthorhombic rare-earth intermetallics. This may be attributable to the competition between ferromagnetic and antiferromagnetic itinerant magnetic exchange interactions and, specifically, competing ordering tendencies at inequivalent Wyckoff sites. The low symmetry of these arrangements of localized magnetic ions among anisotropic conduction bands makes for potentially very complex scenarios. For instance, the positions of Eu ions in the basal plane of \eis~can be viewed as a network of distorted corner-sharing or side-sharing triangles, with 6+1 (in-plane + out-of-plane) distinct nearest-neighbor distances ranging between 3.8\,\AA~and 4.7\,\AA~\cite{SM}. The layer-spacing, 4.6\,\AA, is intermediate to these values, which emphasizes that competition between in-plane and out-of-plane itinerant exchange is likely, and magnetic correlations are not necessarily two-dimensional. Measurements of the magnetic dynamics in these materials would be of great interest to clarify the hierarchy of correlations. 

The small Ising-like net-magnetization of the \G-type magnetic component is a central finding with important consequences. While dipolar interactions are unlikely a driving force of the magnetic reordering at \TZ, the net magnetization must necessarily lift degeneracies of the electronic structure, which was previously not taken into account in density functional calculations~\cite{Rosa2020,Varnava2022}. The hysteretic effects imply that magnetic domains play an important role in \eis, that the domain structure is coupled to charge transport, and that this relation changes qualitatively with temperature. Mesoscale transport studies would now be of interest to clarify whether these observations arise from intrinsic variations of the transport channels or extrinsic mechanisms (scattering of charge carriers from domain walls).

The magnetic phase transition at \TZ~is driven by a competition of ferro- and antiferromagnetic coupling along the $c$ axis. The two possible scenarios, phase-separation or double-$q$ magnetic order, impose different constraints on the spin arrangement but overall have a similar number of degrees of freedom. Since both models are compatible with magnetic scattering, local and bulk magnetic probes~\cite{SM}, their distinction will require direct evidence from spatially resolved techniques. Notably, both possibilities are unusual results: The double-$q$ structure mixes distinct commensurate structures with different ratios at the three Eu sublattices but only provides a physical solution if these components are strictly orthogonal. The concept of phase separation, as illustrated in Fig.~\ref{Fig4}(a,b) is more straightforward, but it is unclear what microscopic mechanism in \eis~could favor inhomogeneity.

\section{Conclusion}
In this study of the magnetism of the non-symmorphic Zintl phase \eis, a combination of local, bulk and momentum-resolved probes of magnetism allowed us to reveal an unusual magnetic ordering process that is also consistent with hysteretic phenomena in both magnetisation and charge transport. This complexity originates in the combination of perpendicular glide symmetries and the distribution of localized magnetic moments on three magnetic sublattices. As a consequence, a complex arrangement of ten co-planar Eu spins is endowed with a net magnetic moment along the $a$ axis, which makes each layer an Ising weak ferromagnet. The demagnetization energy therefore takes a decisive role in the bulk magnetism, and a complex behavior of net magnetized domains may be expected on the mesoscale. At lower temperatures, competing interlayer exchange couplings induce a compensated antiferromagnetic component that displaces the weak ferromagnetism. In the absence of microscopic evidence, it is not possible to distinguish whether this transition occurs in the form of a double-$q$ structure or phase separation.   

It will be of great interest to clarify what bearing the observed magnetic symmetries have on various proposals for crystalline topological phases in the 5-2-6 Zintls~\cite{Wieder2018,Varnava2022}. For instance, phase separation would imply the presence of \Z-type magnetic regimes that conserve effective time-reversal symmetry (time reversal and translation along $c$), a prerequisite for the axion insulator state. In either case, the net magnetization of the (variable) \G~components will necessarily have an impact on the valance bands and might also serve as an unexpected, adjustable (by domain size) parameter of the electronic structure. Furthermore, 5-2-6 Zintls have also been noted for their chemical tunability, and density functional studies predict that the putative topological features in their band structure are extremely sensitive to hydrostatic and uniaxial pressure~\cite{Varnava2022}. Among the wider family of nonsymmorphic orthorhombic intermetallics, there are several structural families in which bulk probes point to similar complex ordering scenarios. Among these are compounds with trivalent rare earth sites that provide similar potential for non-trivial band topology~\cite{Vergniory2022} but open up additional possibilities to tune magnetism. Combining the technological potential of weakly ferromagnetic domains and band topology in such materials is an exciting new avenue to manipulate topological transport and thus to realize specific technological proposals from the emerging field of topological antiferromagnetic spintronics~\cite{Smejkal2018,Varnava2018,Varnava2021}.\\

\newpage
\textbf{Note added.}~\\
During the preparation of this manuscript, we became aware of a single crystal neutron diffraction study by Morano~\etal, Ref.~\cite{Morano2023}. Although the datasets are complementary (in terms of the techniques used and the regime of reciprocal space probed), the type of magnetic structures inferred are consistent with our findings. As in our case, the refinement of the single crystal neutron scattering intensities does not allow a clear distinction between the phase separation and double-$q$ models of the \Z~magnetic component. However, Morano~\etal~observe additional \G-type magnetic Bragg intensities with qualitatively different temperature dependences. This implies that the three Eu sublattices have different ordering tendencies already below \TN -- which adds to the plausibility of the double-$q$ scenario. On the other hand, density functional calculations by Morano~\etal~do favor phase separation. Direct evidence, e.g. from magnetic imaging, will help to clarify this ambiguity.

\begin{center}
{\bf Methods}
\end{center}
\small
\begin{center}
{Samples}
\end{center}
Single crystals of \eis~were prepared using an In-Sb flux method, as previously described~\cite{Rosa2020}. This growth technique yields prismatic rods of \eis~(ca. $1\times0.5\times0.2$\,mm$^3$), along with thinner needles, which we identified as the Eu$_{11}$InSb$_9$ phase~\cite{Fender2021}. To obtain a high-quality powder sample (0.55\,g overall), clean \eis~crystals were selected from a large number of batches, cleaned, and ground. 

\begin{center}
{Neutron powder diffraction}
\end{center}
Time-of-flight neutron powder diffraction (NPD) was performed at WISH (ISIS, Rutherford Appleton Laboratory)~\cite{Chapon2011}, which provides strong signal-to-noise down to very small momentum transfers. This was essential to reveal the magnetic scattering from \eis, given the extreme neutron absorption cross section of Eu, as well as In. To minimize absorption, the sample (0.55\,g) was filled into a double-walled vanadium can and cooled in a lHe cryostat (Oxford Instruments). Data was obtained at 1.5\,K, 10.0\,K and 25.0\,K, and was processed using Mantid~\cite{Arnold2014} and FullProf~\cite{Rodriguez1993}.

\begin{center}
{Muon-spin relaxation}
\end{center}
Zero field $\mu$SR studies of polycrystalline samples were performed on the EMU spectrometer at the STFC-ISIS Neutron and Muon source (pulsed source, higher sensitivity to slow relaxation processes), and the GPS spectrometer at the Swiss Muon Source, Paul Scherrer Institut (continuous source, higher time resolution to resolve static magnetism). Measurements on GPS were performed with the muon spin initially rotated 45$^\circ$ from the incident beam direction and made use of a veto detector to remove the signal from muons that missed the sample. Measurements on EMU were performed with the muon spin anti-parallel to the incident beam direction. For the EMU measurements, the sample was mounted on a thick silver plate that stops all muons that miss the sample, yielding a non-relaxing background signal.

\newpage
\begin{center}
{Resonant elastic x-ray scattering}
\end{center}
Due to the large Eu$^{2+}$ ordered magnetic moment and the strong resonant enhancement at the Eu $L$ absorption edges, hard X-ray resonant elastic scattering (REXS) is highly suitable for magnetic structure determination in Eu compounds. REXS studies were performed at the instruments I16 (DIAMOND Light Source, Rutherford Appleton Laboratory)~\cite{Collins2010} and P09 (DESY, Hamburg)~\cite{Strempfer2013}. In both cases, the photon energy was tuned to the Eu $L_3$ ($2p_\frac32\leftrightarrow5d$) absorption edge (6.970\,keV, $\lambda=1.779$\,\AA). Scattered X-rays were detected either by a position-sensitive CMOS detector (DECTRIS) or an avalanche photodiode in combination with a Cu (220) analyzer crystal. The orientation of the magnetic structure factor vector $\hat{M}(Q)$ was characterized either by azimuthal scans (I16) or by full linear polarization analysis (P09)~\cite{Francoual2013}. \\

\begin{center}
{Magnetometry and transport measurements}
\end{center}
Temperature-dependent magnetic susceptibility and magnetization measurements were performed in a Quantum~Design MPMS3 magnetometer. Resistivity measurements were performed in a Quantum Design PPMS using the standard four-point technique and an AC resistance bridge (Lakeshore~372 with 3708 preamp). The current was applied parallel to the $c$ axis. For the rotation experiments, the sample was mounted on a motor-driven rotation stage. The axis of rotation was parallel to the applied current.

\begin{acknowledgments}
The authors would like to thank Jens Müller (Goethe-Universität Frankfurt) for useful discussions. Parts of this research were carried out at DIAMOND Light Source and at the STFC-ISIS Facility, both institutions of the UK Science and Technology Research Council (STFC). Experiments were also performed at S$\mu$S, Paul Scherrer Institut, Switzerland, as well as under beamtimes I-20190272 and I-20210419 at the synchrotron light source PETRA III at DESY, a member of the Helmholtz Association (HGF). The UK effort was supported by the EPSRC Grant no.\ EP/N032128/1. M.N.W. acknowledges the support of the Natural Sciences and Engineering Research Council of Canada (NSERC). Work at Los Alamos was supported by the U.S. Department of Energy, Office of Basic Energy Sciences, Division of Materials Science and Engineering project “Quantum Fluctuations in Narrow-Band Systems. Work performed at the Paul Scherrer Institute, MJ acknowledges support from the Swiss National Science Foundation (SNSF) [200650]. Work at TU Dresden was supported by the Deutsche Forschungsgemeinschaft through the CRC~1143, the Würzburg-Dresden Cluster of Excellence EXC~2147 (ct.qmat), and through the Emmy~Noether program of the German Science Foundation (project number 501391385). L.S.I.V. was supported by the UK Engineering and Physical Sciences Research Council (EPSRC) (Grants No. EP/N027671/1 and No. EP/N034694/1). M.C.R. is grateful for Fellowships provided by the Laboratory Directed Research \& Development Program at Los Alamos and the Humboldt Foundation. Research data from the UK effort will be made available via Durham Collections. Muon and neutron data collected at STFC-ISIS are available under DOI:10.5286/ISIS.E.RB2010313 and DOI:10.5286/ISIS.E.RB1920539.  
\end{acknowledgments}

\bibliography{Eu526bib}

\begin{thebibliography}{37}%
\makeatletter
\providecommand \@ifxundefined [1]{%
 \@ifx{#1\undefined}
}%
\providecommand \@ifnum [1]{%
 \ifnum #1\expandafter \@firstoftwo
 \else \expandafter \@secondoftwo
 \fi
}%
\providecommand \@ifx [1]{%
 \ifx #1\expandafter \@firstoftwo
 \else \expandafter \@secondoftwo
 \fi
}%
\providecommand \natexlab [1]{#1}%
\providecommand \enquote  [1]{``#1''}%
\providecommand \bibnamefont  [1]{#1}%
\providecommand \bibfnamefont [1]{#1}%
\providecommand \citenamefont [1]{#1}%
\providecommand \href@noop [0]{\@secondoftwo}%
\providecommand \href [0]{\begingroup \@sanitize@url \@href}%
\providecommand \@href[1]{\@@startlink{#1}\@@href}%
\providecommand \@@href[1]{\endgroup#1\@@endlink}%
\providecommand \@sanitize@url [0]{\catcode `\\12\catcode `\$12\catcode
  `\&12\catcode `\#12\catcode `\^12\catcode `\_12\catcode `\%12\relax}%
\providecommand \@@startlink[1]{}%
\providecommand \@@endlink[0]{}%
\providecommand \url  [0]{\begingroup\@sanitize@url \@url }%
\providecommand \@url [1]{\endgroup\@href {#1}{\urlprefix }}%
\providecommand \urlprefix  [0]{URL }%
\providecommand \Eprint [0]{\href }%
\providecommand \doibase [0]{http://dx.doi.org/}%
\providecommand \selectlanguage [0]{\@gobble}%
\providecommand \bibinfo  [0]{\@secondoftwo}%
\providecommand \bibfield  [0]{\@secondoftwo}%
\providecommand \translation [1]{[#1]}%
\providecommand \BibitemOpen [0]{}%
\providecommand \bibitemStop [0]{}%
\providecommand \bibitemNoStop [0]{.\EOS\space}%
\providecommand \EOS [0]{\spacefactor3000\relax}%
\providecommand \BibitemShut  [1]{\csname bibitem#1\endcsname}%
\let\auto@bib@innerbib\@empty
\bibitem [{\citenamefont {Bradlyn}\ \emph {et~al.}(2017)\citenamefont
  {Bradlyn}, \citenamefont {Elcoro}, \citenamefont {Cano}, \citenamefont
  {Vergniory}, \citenamefont {Wang}, \citenamefont {Felser}, \citenamefont
  {Aroyo},\ and\ \citenamefont {Bernevig}}]{Bradlyn2017}%
  \BibitemOpen
  \bibfield  {author} {\bibinfo {author} {\bibfnamefont {B.}~\bibnamefont
  {Bradlyn}}, \bibinfo {author} {\bibfnamefont {L.}~\bibnamefont {Elcoro}},
  \bibinfo {author} {\bibfnamefont {J.}~\bibnamefont {Cano}}, \bibinfo {author}
  {\bibfnamefont {M.~G.}\ \bibnamefont {Vergniory}}, \bibinfo {author}
  {\bibfnamefont {Z.}~\bibnamefont {Wang}}, \bibinfo {author} {\bibfnamefont
  {C.}~\bibnamefont {Felser}}, \bibinfo {author} {\bibfnamefont {M.~I.}\
  \bibnamefont {Aroyo}}, \ and\ \bibinfo {author} {\bibfnamefont {B.~A.}\
  \bibnamefont {Bernevig}},\ }\href {\doibase 10.1038/nature23268} {\bibfield
  {journal} {\bibinfo  {journal} {Nature}\ }\textbf {\bibinfo {volume} {547}},\
  \bibinfo {pages} {298} (\bibinfo {year} {2017})}\BibitemShut {NoStop}%
\bibitem [{\citenamefont {Zhang}\ \emph {et~al.}(2019)\citenamefont {Zhang},
  \citenamefont {Jiang}, \citenamefont {Song}, \citenamefont {Huang},
  \citenamefont {He}, \citenamefont {Fang}, \citenamefont {Weng},\ and\
  \citenamefont {Fang}}]{Zhang2019}%
  \BibitemOpen
  \bibfield  {author} {\bibinfo {author} {\bibfnamefont {T.}~\bibnamefont
  {Zhang}}, \bibinfo {author} {\bibfnamefont {Y.}~\bibnamefont {Jiang}},
  \bibinfo {author} {\bibfnamefont {Z.}~\bibnamefont {Song}}, \bibinfo {author}
  {\bibfnamefont {H.}~\bibnamefont {Huang}}, \bibinfo {author} {\bibfnamefont
  {Y.}~\bibnamefont {He}}, \bibinfo {author} {\bibfnamefont {Z.}~\bibnamefont
  {Fang}}, \bibinfo {author} {\bibfnamefont {H.}~\bibnamefont {Weng}}, \ and\
  \bibinfo {author} {\bibfnamefont {C.}~\bibnamefont {Fang}},\ }\href {\doibase
  10.1038/s41586-019-0944-6} {\bibfield  {journal} {\bibinfo  {journal}
  {Nature}\ }\textbf {\bibinfo {volume} {566}},\ \bibinfo {pages} {475}
  (\bibinfo {year} {2019})}\BibitemShut {NoStop}%
\bibitem [{\citenamefont {Vergniory}\ \emph {et~al.}(2019)\citenamefont
  {Vergniory}, \citenamefont {Elcoro}, \citenamefont {Felser}, \citenamefont
  {Regnault}, \citenamefont {Bernevig},\ and\ \citenamefont
  {Wang}}]{Vergniory2019}%
  \BibitemOpen
  \bibfield  {author} {\bibinfo {author} {\bibfnamefont {M.~G.}\ \bibnamefont
  {Vergniory}}, \bibinfo {author} {\bibfnamefont {L.}~\bibnamefont {Elcoro}},
  \bibinfo {author} {\bibfnamefont {C.}~\bibnamefont {Felser}}, \bibinfo
  {author} {\bibfnamefont {N.}~\bibnamefont {Regnault}}, \bibinfo {author}
  {\bibfnamefont {B.~A.}\ \bibnamefont {Bernevig}}, \ and\ \bibinfo {author}
  {\bibfnamefont {Z.}~\bibnamefont {Wang}},\ }\href {\doibase
  10.1038/s41586-019-0954-4} {\bibfield  {journal} {\bibinfo  {journal}
  {Nature}\ }\textbf {\bibinfo {volume} {566}},\ \bibinfo {pages} {480}
  (\bibinfo {year} {2019})}\BibitemShut {NoStop}%
\bibitem [{\citenamefont {Watanabe}\ \emph {et~al.}(2018)\citenamefont
  {Watanabe}, \citenamefont {Po},\ and\ \citenamefont
  {Vishwanath}}]{Watanabe2018}%
  \BibitemOpen
  \bibfield  {author} {\bibinfo {author} {\bibfnamefont {H.}~\bibnamefont
  {Watanabe}}, \bibinfo {author} {\bibfnamefont {H.~C.}\ \bibnamefont {Po}}, \
  and\ \bibinfo {author} {\bibfnamefont {A.}~\bibnamefont {Vishwanath}},\
  }\href {\doibase 10.1126/sciadv.aat8685} {\bibfield  {journal} {\bibinfo
  {journal} {Science Advances}\ }\textbf {\bibinfo {volume} {4}},\ \bibinfo
  {pages} {eaat8685} (\bibinfo {year} {2018})},\ \Eprint
  {http://arxiv.org/abs/https://www.science.org/doi/pdf/10.1126/sciadv.aat8685}
  {https://www.science.org/doi/pdf/10.1126/sciadv.aat8685} \BibitemShut
  {NoStop}%
\bibitem [{\citenamefont {Xu}\ \emph {et~al.}(2020)\citenamefont {Xu},
  \citenamefont {Elcoro}, \citenamefont {Song}, \citenamefont {Wieder},
  \citenamefont {Vergniory}, \citenamefont {Regnault}, \citenamefont {Chen},
  \citenamefont {Felser},\ and\ \citenamefont {Bernevig}}]{Xu2020}%
  \BibitemOpen
  \bibfield  {author} {\bibinfo {author} {\bibfnamefont {Y.}~\bibnamefont
  {Xu}}, \bibinfo {author} {\bibfnamefont {L.}~\bibnamefont {Elcoro}}, \bibinfo
  {author} {\bibfnamefont {Z.-D.}\ \bibnamefont {Song}}, \bibinfo {author}
  {\bibfnamefont {B.~J.}\ \bibnamefont {Wieder}}, \bibinfo {author}
  {\bibfnamefont {M.~G.}\ \bibnamefont {Vergniory}}, \bibinfo {author}
  {\bibfnamefont {N.}~\bibnamefont {Regnault}}, \bibinfo {author}
  {\bibfnamefont {Y.}~\bibnamefont {Chen}}, \bibinfo {author} {\bibfnamefont
  {C.}~\bibnamefont {Felser}}, \ and\ \bibinfo {author} {\bibfnamefont {B.~A.}\
  \bibnamefont {Bernevig}},\ }\href {\doibase 10.1038/s41586-020-2837-0}
  {\bibfield  {journal} {\bibinfo  {journal} {Nature}\ }\textbf {\bibinfo
  {volume} {586}},\ \bibinfo {pages} {702} (\bibinfo {year}
  {2020})}\BibitemShut {NoStop}%
\bibitem [{\citenamefont {Wieder}\ \emph {et~al.}(2018)\citenamefont {Wieder},
  \citenamefont {Bradlyn}, \citenamefont {Wang}, \citenamefont {Cano},
  \citenamefont {Kim}, \citenamefont {Kim}, \citenamefont {Rappe},
  \citenamefont {Kane},\ and\ \citenamefont {Bernevig}}]{Wieder2018}%
  \BibitemOpen
  \bibfield  {author} {\bibinfo {author} {\bibfnamefont {B.~J.}\ \bibnamefont
  {Wieder}}, \bibinfo {author} {\bibfnamefont {B.}~\bibnamefont {Bradlyn}},
  \bibinfo {author} {\bibfnamefont {Z.}~\bibnamefont {Wang}}, \bibinfo {author}
  {\bibfnamefont {J.}~\bibnamefont {Cano}}, \bibinfo {author} {\bibfnamefont
  {Y.}~\bibnamefont {Kim}}, \bibinfo {author} {\bibfnamefont {H.-S.~D.}\
  \bibnamefont {Kim}}, \bibinfo {author} {\bibfnamefont {A.~M.}\ \bibnamefont
  {Rappe}}, \bibinfo {author} {\bibfnamefont {C.~L.}\ \bibnamefont {Kane}}, \
  and\ \bibinfo {author} {\bibfnamefont {B.~A.}\ \bibnamefont {Bernevig}},\
  }\href {\doibase 10.1126/science.aan2802} {\bibfield  {journal} {\bibinfo
  {journal} {Science}\ }\textbf {\bibinfo {volume} {361}},\ \bibinfo {pages}
  {246} (\bibinfo {year} {2018})},\ \Eprint
  {http://arxiv.org/abs/https://www.science.org/doi/pdf/10.1126/science.aan2802}
  {https://www.science.org/doi/pdf/10.1126/science.aan2802} \BibitemShut
  {NoStop}%
\bibitem [{\citenamefont {Chanakian}\ \emph {et~al.}(2015)\citenamefont
  {Chanakian}, \citenamefont {Aydemir}, \citenamefont {Zevalkink},
  \citenamefont {Gibbs}, \citenamefont {Fleurial}, \citenamefont {Bux},\ and\
  \citenamefont {Jeffrey~Snyder}}]{Chanakian2015}%
  \BibitemOpen
  \bibfield  {author} {\bibinfo {author} {\bibfnamefont {S.}~\bibnamefont
  {Chanakian}}, \bibinfo {author} {\bibfnamefont {U.}~\bibnamefont {Aydemir}},
  \bibinfo {author} {\bibfnamefont {A.}~\bibnamefont {Zevalkink}}, \bibinfo
  {author} {\bibfnamefont {Z.~M.}\ \bibnamefont {Gibbs}}, \bibinfo {author}
  {\bibfnamefont {J.-P.}\ \bibnamefont {Fleurial}}, \bibinfo {author}
  {\bibfnamefont {S.}~\bibnamefont {Bux}}, \ and\ \bibinfo {author}
  {\bibfnamefont {G.}~\bibnamefont {Jeffrey~Snyder}},\ }\href {\doibase
  10.1039/C5TC01645B} {\bibfield  {journal} {\bibinfo  {journal} {J. Mater.
  Chem. C}\ }\textbf {\bibinfo {volume} {3}},\ \bibinfo {pages} {10518}
  (\bibinfo {year} {2015})}\BibitemShut {NoStop}%
\bibitem [{\citenamefont {Lv}\ \emph {et~al.}(2017)\citenamefont {Lv},
  \citenamefont {Yang}, \citenamefont {Lin}, \citenamefont {Hu}, \citenamefont
  {Guo}, \citenamefont {Yang}, \citenamefont {Luo},\ and\ \citenamefont
  {Zhao}}]{Lv2017}%
  \BibitemOpen
  \bibfield  {author} {\bibinfo {author} {\bibfnamefont {W.}~\bibnamefont
  {Lv}}, \bibinfo {author} {\bibfnamefont {C.}~\bibnamefont {Yang}}, \bibinfo
  {author} {\bibfnamefont {J.}~\bibnamefont {Lin}}, \bibinfo {author}
  {\bibfnamefont {X.}~\bibnamefont {Hu}}, \bibinfo {author} {\bibfnamefont
  {K.}~\bibnamefont {Guo}}, \bibinfo {author} {\bibfnamefont {X.}~\bibnamefont
  {Yang}}, \bibinfo {author} {\bibfnamefont {J.}~\bibnamefont {Luo}}, \ and\
  \bibinfo {author} {\bibfnamefont {J.-T.}\ \bibnamefont {Zhao}},\ }\href
  {\doibase https://doi.org/10.1016/j.jallcom.2017.08.033} {\bibfield
  {journal} {\bibinfo  {journal} {Journal of Alloys and Compounds}\ }\textbf
  {\bibinfo {volume} {726}},\ \bibinfo {pages} {618} (\bibinfo {year}
  {2017})}\BibitemShut {NoStop}%
\bibitem [{\citenamefont {Rosa}\ \emph {et~al.}(2020)\citenamefont {Rosa},
  \citenamefont {Xu}, \citenamefont {Rahn}, \citenamefont {Souza},
  \citenamefont {Kushwaha}, \citenamefont {Veiga}, \citenamefont {Bombardi},
  \citenamefont {Thomas}, \citenamefont {Janoschek}, \citenamefont {Bauer},
  \citenamefont {Chan}, \citenamefont {Wang}, \citenamefont {Thompson},
  \citenamefont {Harrison}, \citenamefont {Pagliuso}, \citenamefont
  {Bernevig},\ and\ \citenamefont {Ronning}}]{Rosa2020}%
  \BibitemOpen
  \bibfield  {author} {\bibinfo {author} {\bibfnamefont {P.~F.~S.}\
  \bibnamefont {Rosa}}, \bibinfo {author} {\bibfnamefont {Y.}~\bibnamefont
  {Xu}}, \bibinfo {author} {\bibfnamefont {M.}~\bibnamefont {Rahn}}, \bibinfo
  {author} {\bibfnamefont {J.}~\bibnamefont {Souza}}, \bibinfo {author}
  {\bibfnamefont {S.}~\bibnamefont {Kushwaha}}, \bibinfo {author}
  {\bibfnamefont {L.}~\bibnamefont {Veiga}}, \bibinfo {author} {\bibfnamefont
  {A.}~\bibnamefont {Bombardi}}, \bibinfo {author} {\bibfnamefont
  {S.}~\bibnamefont {Thomas}}, \bibinfo {author} {\bibfnamefont
  {M.}~\bibnamefont {Janoschek}}, \bibinfo {author} {\bibfnamefont {E.~D.}\
  \bibnamefont {Bauer}}, \bibinfo {author} {\bibfnamefont {M.}~\bibnamefont
  {Chan}}, \bibinfo {author} {\bibfnamefont {Z.}~\bibnamefont {Wang}}, \bibinfo
  {author} {\bibfnamefont {J.}~\bibnamefont {Thompson}}, \bibinfo {author}
  {\bibfnamefont {N.}~\bibnamefont {Harrison}}, \bibinfo {author}
  {\bibfnamefont {P.}~\bibnamefont {Pagliuso}}, \bibinfo {author}
  {\bibfnamefont {A.}~\bibnamefont {Bernevig}}, \ and\ \bibinfo {author}
  {\bibfnamefont {F.}~\bibnamefont {Ronning}},\ }\href {\doibase
  10.1038/s41535-020-00256-8} {\bibfield  {journal} {\bibinfo  {journal} {npj
  Quantum Materials}\ }\textbf {\bibinfo {volume} {5}},\ \bibinfo {pages} {52}
  (\bibinfo {year} {2020})}\BibitemShut {NoStop}%
\bibitem [{\citenamefont {Souza}\ \emph {et~al.}(2022)\citenamefont {Souza},
  \citenamefont {Thomas}, \citenamefont {Bauer}, \citenamefont {Thompson},
  \citenamefont {Ronning}, \citenamefont {Pagliuso},\ and\ \citenamefont
  {Rosa}}]{Souza2022}%
  \BibitemOpen
  \bibfield  {author} {\bibinfo {author} {\bibfnamefont {J.~C.}\ \bibnamefont
  {Souza}}, \bibinfo {author} {\bibfnamefont {S.~M.}\ \bibnamefont {Thomas}},
  \bibinfo {author} {\bibfnamefont {E.~D.}\ \bibnamefont {Bauer}}, \bibinfo
  {author} {\bibfnamefont {J.~D.}\ \bibnamefont {Thompson}}, \bibinfo {author}
  {\bibfnamefont {F.}~\bibnamefont {Ronning}}, \bibinfo {author} {\bibfnamefont
  {P.~G.}\ \bibnamefont {Pagliuso}}, \ and\ \bibinfo {author} {\bibfnamefont
  {P.~F.~S.}\ \bibnamefont {Rosa}},\ }\href {\doibase
  10.1103/PhysRevB.105.035135} {\bibfield  {journal} {\bibinfo  {journal}
  {Phys. Rev. B}\ }\textbf {\bibinfo {volume} {105}},\ \bibinfo {pages}
  {035135} (\bibinfo {year} {2022})}\BibitemShut {NoStop}%
\bibitem [{\citenamefont {Ghosh}\ \emph {et~al.}(2022)\citenamefont {Ghosh},
  \citenamefont {Lane}, \citenamefont {Ronning}, \citenamefont {Bauer},
  \citenamefont {Thompson}, \citenamefont {Zhu}, \citenamefont {Rosa},\ and\
  \citenamefont {Thomas}}]{Ghosh2022}%
  \BibitemOpen
  \bibfield  {author} {\bibinfo {author} {\bibfnamefont {S.}~\bibnamefont
  {Ghosh}}, \bibinfo {author} {\bibfnamefont {C.}~\bibnamefont {Lane}},
  \bibinfo {author} {\bibfnamefont {F.}~\bibnamefont {Ronning}}, \bibinfo
  {author} {\bibfnamefont {E.~D.}\ \bibnamefont {Bauer}}, \bibinfo {author}
  {\bibfnamefont {J.~D.}\ \bibnamefont {Thompson}}, \bibinfo {author}
  {\bibfnamefont {J.-X.}\ \bibnamefont {Zhu}}, \bibinfo {author} {\bibfnamefont
  {P.~F.~S.}\ \bibnamefont {Rosa}}, \ and\ \bibinfo {author} {\bibfnamefont
  {S.~M.}\ \bibnamefont {Thomas}},\ }\href {\doibase
  10.1103/PhysRevB.106.045110} {\bibfield  {journal} {\bibinfo  {journal}
  {Phys. Rev. B}\ }\textbf {\bibinfo {volume} {106}},\ \bibinfo {pages}
  {045110} (\bibinfo {year} {2022})}\BibitemShut {NoStop}%
\bibitem [{\citenamefont {Varnava}\ \emph {et~al.}(2022)\citenamefont
  {Varnava}, \citenamefont {Berry}, \citenamefont {McQueen},\ and\
  \citenamefont {Vanderbilt}}]{Varnava2022}%
  \BibitemOpen
  \bibfield  {author} {\bibinfo {author} {\bibfnamefont {N.}~\bibnamefont
  {Varnava}}, \bibinfo {author} {\bibfnamefont {T.}~\bibnamefont {Berry}},
  \bibinfo {author} {\bibfnamefont {T.~M.}\ \bibnamefont {McQueen}}, \ and\
  \bibinfo {author} {\bibfnamefont {D.}~\bibnamefont {Vanderbilt}},\ }\href
  {\doibase 10.1103/PhysRevB.105.235128} {\bibfield  {journal} {\bibinfo
  {journal} {Phys. Rev. B}\ }\textbf {\bibinfo {volume} {105}},\ \bibinfo
  {pages} {235128} (\bibinfo {year} {2022})}\BibitemShut {NoStop}%
\bibitem [{SMF()}]{SMFigS3}%
  \BibitemOpen
  \href@noop {} {\enquote {\bibinfo {title} {See {Fig.~S3} in {Supplementary
  Note 1} of the {Supplemental Material} attached to this preprint},}\
  }\BibitemShut {NoStop}%
\bibitem [{\citenamefont {Sekine}\ and\ \citenamefont
  {Nomura}(2021)}]{Sekine2021}%
  \BibitemOpen
  \bibfield  {author} {\bibinfo {author} {\bibfnamefont {A.}~\bibnamefont
  {Sekine}}\ and\ \bibinfo {author} {\bibfnamefont {K.}~\bibnamefont
  {Nomura}},\ }\href {\doibase 10.1063/5.0038804} {\bibfield  {journal}
  {\bibinfo  {journal} {Journal of Applied Physics}\ }\textbf {\bibinfo
  {volume} {129}},\ \bibinfo {pages} {141101} (\bibinfo {year} {2021})},\
  \Eprint {http://arxiv.org/abs/https://doi.org/10.1063/5.0038804}
  {https://doi.org/10.1063/5.0038804} \BibitemShut {NoStop}%
\bibitem [{\citenamefont {Bernevig}\ \emph {et~al.}(2022)\citenamefont
  {Bernevig}, \citenamefont {Felser},\ and\ \citenamefont
  {Beidenkopf}}]{Bernevig2022}%
  \BibitemOpen
  \bibfield  {author} {\bibinfo {author} {\bibfnamefont {B.~A.}\ \bibnamefont
  {Bernevig}}, \bibinfo {author} {\bibfnamefont {C.}~\bibnamefont {Felser}}, \
  and\ \bibinfo {author} {\bibfnamefont {H.}~\bibnamefont {Beidenkopf}},\
  }\href {\doibase 10.1038/s41586-021-04105-x} {\bibfield  {journal} {\bibinfo
  {journal} {Nature}\ }\textbf {\bibinfo {volume} {603}},\ \bibinfo {pages}
  {41} (\bibinfo {year} {2022})}\BibitemShut {NoStop}%
\bibitem [{SN4()}]{SN4}%
  \BibitemOpen
  \href@noop {} {\enquote {\bibinfo {title} {Details of the neutron scattering
  experiment and data analysis are discussed in {Supplementary Notes 3 and 4}
  of the {Supplemental Material} attached to this preprint},}\ }\BibitemShut
  {NoStop}%
\bibitem [{SM()}]{SM}%
  \BibitemOpen
  \href@noop {} {\enquote {\bibinfo {title} {{Supplemental Material} attached
  to this preprint},}\ }\BibitemShut {NoStop}%
\bibitem [{SN1()}]{SN1SN2}%
  \BibitemOpen
  \href@noop {} {\enquote {\bibinfo {title} {Detailed information regarding
  representational analysis and magnetic structure factors in {Supplementary
  Notes 1 and 2} of the {Supplemental Material} attached to this preprint},}\
  }\BibitemShut {NoStop}%
\bibitem [{\citenamefont {Hicken}\ \emph {et~al.}(2020)\citenamefont {Hicken},
  \citenamefont {Holt}, \citenamefont {Franke}, \citenamefont {Hawkhead},
  \citenamefont {\ifmmode \check{S}\else \v{S}\fi{}tefan\ifmmode \check{c}\else
  \v{c}\fi{}i\ifmmode~\check{c}\else \v{c}\fi{}}, \citenamefont {Wilson},
  \citenamefont {Gomil\ifmmode~\check{s}\else \v{s}\fi{}ek}, \citenamefont
  {Huddart}, \citenamefont {Clark}, \citenamefont {Lees}, \citenamefont
  {Pratt}, \citenamefont {Blundell}, \citenamefont {Balakrishnan},\ and\
  \citenamefont {Lancaster}}]{Hicken2020}%
  \BibitemOpen
  \bibfield  {author} {\bibinfo {author} {\bibfnamefont {T.~J.}\ \bibnamefont
  {Hicken}}, \bibinfo {author} {\bibfnamefont {S.~J.~R.}\ \bibnamefont {Holt}},
  \bibinfo {author} {\bibfnamefont {K.~J.~A.}\ \bibnamefont {Franke}}, \bibinfo
  {author} {\bibfnamefont {Z.}~\bibnamefont {Hawkhead}}, \bibinfo {author}
  {\bibfnamefont {A.}~\bibnamefont {\ifmmode \check{S}\else
  \v{S}\fi{}tefan\ifmmode \check{c}\else \v{c}\fi{}i\ifmmode~\check{c}\else
  \v{c}\fi{}}}, \bibinfo {author} {\bibfnamefont {M.~N.}\ \bibnamefont
  {Wilson}}, \bibinfo {author} {\bibfnamefont {M.}~\bibnamefont
  {Gomil\ifmmode~\check{s}\else \v{s}\fi{}ek}}, \bibinfo {author}
  {\bibfnamefont {B.~M.}\ \bibnamefont {Huddart}}, \bibinfo {author}
  {\bibfnamefont {S.~J.}\ \bibnamefont {Clark}}, \bibinfo {author}
  {\bibfnamefont {M.~R.}\ \bibnamefont {Lees}}, \bibinfo {author}
  {\bibfnamefont {F.~L.}\ \bibnamefont {Pratt}}, \bibinfo {author}
  {\bibfnamefont {S.~J.}\ \bibnamefont {Blundell}}, \bibinfo {author}
  {\bibfnamefont {G.}~\bibnamefont {Balakrishnan}}, \ and\ \bibinfo {author}
  {\bibfnamefont {T.}~\bibnamefont {Lancaster}},\ }\href {\doibase
  10.1103/PhysRevResearch.2.032001} {\bibfield  {journal} {\bibinfo  {journal}
  {Phys. Rev. Res.}\ }\textbf {\bibinfo {volume} {2}},\ \bibinfo {pages}
  {032001} (\bibinfo {year} {2020})}\BibitemShut {NoStop}%
\bibitem [{\citenamefont {Franke}\ \emph {et~al.}(2018)\citenamefont {Franke},
  \citenamefont {Huddart}, \citenamefont {Hicken}, \citenamefont {Xiao},
  \citenamefont {Blundell}, \citenamefont {Pratt}, \citenamefont {Crisanti},
  \citenamefont {Barker}, \citenamefont {Clark}, \citenamefont {\ifmmode
  \check{S}\else \v{S}\fi{}tefan\ifmmode \check{c}\else
  \v{c}\fi{}i\ifmmode~\check{c}\else \v{c}\fi{}}, \citenamefont {Hatnean},
  \citenamefont {Balakrishnan},\ and\ \citenamefont {Lancaster}}]{Franke2018}%
  \BibitemOpen
  \bibfield  {author} {\bibinfo {author} {\bibfnamefont {K.~J.~A.}\
  \bibnamefont {Franke}}, \bibinfo {author} {\bibfnamefont {B.~M.}\
  \bibnamefont {Huddart}}, \bibinfo {author} {\bibfnamefont {T.~J.}\
  \bibnamefont {Hicken}}, \bibinfo {author} {\bibfnamefont {F.}~\bibnamefont
  {Xiao}}, \bibinfo {author} {\bibfnamefont {S.~J.}\ \bibnamefont {Blundell}},
  \bibinfo {author} {\bibfnamefont {F.~L.}\ \bibnamefont {Pratt}}, \bibinfo
  {author} {\bibfnamefont {M.}~\bibnamefont {Crisanti}}, \bibinfo {author}
  {\bibfnamefont {J.~A.~T.}\ \bibnamefont {Barker}}, \bibinfo {author}
  {\bibfnamefont {S.~J.}\ \bibnamefont {Clark}}, \bibinfo {author}
  {\bibfnamefont {A.~c.~v.}\ \bibnamefont {\ifmmode \check{S}\else
  \v{S}\fi{}tefan\ifmmode \check{c}\else \v{c}\fi{}i\ifmmode~\check{c}\else
  \v{c}\fi{}}}, \bibinfo {author} {\bibfnamefont {M.~C.}\ \bibnamefont
  {Hatnean}}, \bibinfo {author} {\bibfnamefont {G.}~\bibnamefont
  {Balakrishnan}}, \ and\ \bibinfo {author} {\bibfnamefont {T.}~\bibnamefont
  {Lancaster}},\ }\href {\doibase 10.1103/PhysRevB.98.054428} {\bibfield
  {journal} {\bibinfo  {journal} {Phys. Rev. B}\ }\textbf {\bibinfo {volume}
  {98}},\ \bibinfo {pages} {054428} (\bibinfo {year} {2018})}\BibitemShut
  {NoStop}%
\bibitem [{\citenamefont {Guo}\ \emph {et~al.}(2022)\citenamefont {Guo},
  \citenamefont {Wang}, \citenamefont {Wang}, \citenamefont {Gu}, \citenamefont
  {Mi}, \citenamefont {Zhu}, \citenamefont {Hu}, \citenamefont {Pang},
  \citenamefont {Ji}, \citenamefont {Gao}, \citenamefont {Xia},\ and\
  \citenamefont {Cheng}}]{Guo2022}%
  \BibitemOpen
  \bibfield  {author} {\bibinfo {author} {\bibfnamefont {J.}~\bibnamefont
  {Guo}}, \bibinfo {author} {\bibfnamefont {H.}~\bibnamefont {Wang}}, \bibinfo
  {author} {\bibfnamefont {X.}~\bibnamefont {Wang}}, \bibinfo {author}
  {\bibfnamefont {S.}~\bibnamefont {Gu}}, \bibinfo {author} {\bibfnamefont
  {S.}~\bibnamefont {Mi}}, \bibinfo {author} {\bibfnamefont {S.}~\bibnamefont
  {Zhu}}, \bibinfo {author} {\bibfnamefont {J.}~\bibnamefont {Hu}}, \bibinfo
  {author} {\bibfnamefont {F.}~\bibnamefont {Pang}}, \bibinfo {author}
  {\bibfnamefont {W.}~\bibnamefont {Ji}}, \bibinfo {author} {\bibfnamefont
  {H.-J.}\ \bibnamefont {Gao}}, \bibinfo {author} {\bibfnamefont
  {T.}~\bibnamefont {Xia}}, \ and\ \bibinfo {author} {\bibfnamefont
  {Z.}~\bibnamefont {Cheng}},\ }\href {\doibase 10.1021/acs.jpcc.2c02223}
  {\bibfield  {journal} {\bibinfo  {journal} {The Journal of Physical Chemistry
  C}\ }\textbf {\bibinfo {volume} {126}},\ \bibinfo {pages} {13884} (\bibinfo
  {year} {2022})}\BibitemShut {NoStop}%
\bibitem [{\citenamefont {Papanicolaou}(1995)}]{Papanicolaou1995}%
  \BibitemOpen
  \bibfield  {author} {\bibinfo {author} {\bibfnamefont {N.}~\bibnamefont
  {Papanicolaou}},\ }\href {\doibase 10.1103/PhysRevB.51.15062} {\bibfield
  {journal} {\bibinfo  {journal} {Phys. Rev. B}\ }\textbf {\bibinfo {volume}
  {51}},\ \bibinfo {pages} {15062} (\bibinfo {year} {1995})}\BibitemShut
  {NoStop}%
\bibitem [{\citenamefont {Szytuła}\ \emph {et~al.}(2014)\citenamefont
  {Szytuła}, \citenamefont {Baran}, \citenamefont {Kaczorowski}, \citenamefont
  {Sikora},\ and\ \citenamefont {Hoser}}]{Szytula2014}%
  \BibitemOpen
  \bibfield  {author} {\bibinfo {author} {\bibfnamefont {A.}~\bibnamefont
  {Szytuła}}, \bibinfo {author} {\bibfnamefont {S.}~\bibnamefont {Baran}},
  \bibinfo {author} {\bibfnamefont {D.}~\bibnamefont {Kaczorowski}}, \bibinfo
  {author} {\bibfnamefont {W.}~\bibnamefont {Sikora}}, \ and\ \bibinfo {author}
  {\bibfnamefont {A.}~\bibnamefont {Hoser}},\ }\href {\doibase
  https://doi.org/10.1016/j.jallcom.2014.07.190} {\bibfield  {journal}
  {\bibinfo  {journal} {Journal of Alloys and Compounds}\ }\textbf {\bibinfo
  {volume} {617}},\ \bibinfo {pages} {149} (\bibinfo {year}
  {2014})}\BibitemShut {NoStop}%
\bibitem [{\citenamefont {Provino}\ \emph {et~al.}(2012)\citenamefont
  {Provino}, \citenamefont {Mudryk}, \citenamefont {Paudyal}, \citenamefont
  {Smetana}, \citenamefont {Manfrinetti}, \citenamefont {Pecharsky},
  \citenamefont {Gschneidner},\ and\ \citenamefont {Corbett}}]{Provino2012}%
  \BibitemOpen
  \bibfield  {author} {\bibinfo {author} {\bibfnamefont {A.}~\bibnamefont
  {Provino}}, \bibinfo {author} {\bibfnamefont {Y.}~\bibnamefont {Mudryk}},
  \bibinfo {author} {\bibfnamefont {D.}~\bibnamefont {Paudyal}}, \bibinfo
  {author} {\bibfnamefont {V.}~\bibnamefont {Smetana}}, \bibinfo {author}
  {\bibfnamefont {P.}~\bibnamefont {Manfrinetti}}, \bibinfo {author}
  {\bibfnamefont {V.~K.}\ \bibnamefont {Pecharsky}}, \bibinfo {author}
  {\bibfnamefont {J.}~\bibnamefont {Gschneidner}, \bibfnamefont {K.~A.}}, \
  and\ \bibinfo {author} {\bibfnamefont {J.~D.}\ \bibnamefont {Corbett}},\
  }\href {\doibase 10.1063/1.3673432} {\bibfield  {journal} {\bibinfo
  {journal} {Journal of Applied Physics}\ }\textbf {\bibinfo {volume} {111}},\
  \bibinfo {pages} {07E122} (\bibinfo {year} {2012})},\ \Eprint
  {http://arxiv.org/abs/https://pubs.aip.org/aip/jap/article-pdf/doi/10.1063/1.3673432/14894538/07e122\_1\_online.pdf}
  {https://pubs.aip.org/aip/jap/article-pdf/doi/10.1063/1.3673432/14894538/07e122\_1\_online.pdf}
  \BibitemShut {NoStop}%
\bibitem [{SN9()}]{SN9}%
  \BibitemOpen
  \href@noop {} {\enquote {\bibinfo {title} {Illustration of nearest neighbor
  exchange paths in {Supplementary Note 10} of the {Supplemental Material}
  attached to this preprint},}\ }\BibitemShut {NoStop}%
\bibitem [{\citenamefont {Vergniory}\ \emph {et~al.}(2022)\citenamefont
  {Vergniory}, \citenamefont {Wieder}, \citenamefont {Elcoro}, \citenamefont
  {Parkin}, \citenamefont {Felser}, \citenamefont {Bernevig},\ and\
  \citenamefont {Regnault}}]{Vergniory2022}%
  \BibitemOpen
  \bibfield  {author} {\bibinfo {author} {\bibfnamefont {M.~G.}\ \bibnamefont
  {Vergniory}}, \bibinfo {author} {\bibfnamefont {B.~J.}\ \bibnamefont
  {Wieder}}, \bibinfo {author} {\bibfnamefont {L.}~\bibnamefont {Elcoro}},
  \bibinfo {author} {\bibfnamefont {S.~S.~P.}\ \bibnamefont {Parkin}}, \bibinfo
  {author} {\bibfnamefont {C.}~\bibnamefont {Felser}}, \bibinfo {author}
  {\bibfnamefont {B.~A.}\ \bibnamefont {Bernevig}}, \ and\ \bibinfo {author}
  {\bibfnamefont {N.}~\bibnamefont {Regnault}},\ }\href {\doibase
  10.1126/science.abg9094} {\bibfield  {journal} {\bibinfo  {journal}
  {Science}\ }\textbf {\bibinfo {volume} {376}},\ \bibinfo {pages} {eabg9094}
  (\bibinfo {year} {2022})},\ \Eprint
  {http://arxiv.org/abs/https://www.science.org/doi/pdf/10.1126/science.abg9094}
  {https://www.science.org/doi/pdf/10.1126/science.abg9094} \BibitemShut
  {NoStop}%
\bibitem [{\citenamefont {{\v S}mejkal}\ \emph {et~al.}(2018)\citenamefont {{\v
  S}mejkal}, \citenamefont {Mokrousov}, \citenamefont {Yan},\ and\
  \citenamefont {MacDonald}}]{Smejkal2018}%
  \BibitemOpen
  \bibfield  {author} {\bibinfo {author} {\bibfnamefont {L.}~\bibnamefont {{\v
  S}mejkal}}, \bibinfo {author} {\bibfnamefont {Y.}~\bibnamefont {Mokrousov}},
  \bibinfo {author} {\bibfnamefont {B.}~\bibnamefont {Yan}}, \ and\ \bibinfo
  {author} {\bibfnamefont {A.~H.}\ \bibnamefont {MacDonald}},\ }\href {\doibase
  10.1038/s41567-018-0064-5} {\bibfield  {journal} {\bibinfo  {journal} {Nature
  Physics}\ }\textbf {\bibinfo {volume} {14}},\ \bibinfo {pages} {242}
  (\bibinfo {year} {2018})}\BibitemShut {NoStop}%
\bibitem [{\citenamefont {Varnava}\ and\ \citenamefont
  {Vanderbilt}(2018)}]{Varnava2018}%
  \BibitemOpen
  \bibfield  {author} {\bibinfo {author} {\bibfnamefont {N.}~\bibnamefont
  {Varnava}}\ and\ \bibinfo {author} {\bibfnamefont {D.}~\bibnamefont
  {Vanderbilt}},\ }\href {\doibase 10.1103/PhysRevB.98.245117} {\bibfield
  {journal} {\bibinfo  {journal} {Phys. Rev. B}\ }\textbf {\bibinfo {volume}
  {98}},\ \bibinfo {pages} {245117} (\bibinfo {year} {2018})}\BibitemShut
  {NoStop}%
\bibitem [{\citenamefont {Varnava}\ \emph {et~al.}(2021)\citenamefont
  {Varnava}, \citenamefont {Wilson}, \citenamefont {Pixley},\ and\
  \citenamefont {Vanderbilt}}]{Varnava2021}%
  \BibitemOpen
  \bibfield  {author} {\bibinfo {author} {\bibfnamefont {N.}~\bibnamefont
  {Varnava}}, \bibinfo {author} {\bibfnamefont {J.~H.}\ \bibnamefont {Wilson}},
  \bibinfo {author} {\bibfnamefont {J.~H.}\ \bibnamefont {Pixley}}, \ and\
  \bibinfo {author} {\bibfnamefont {D.}~\bibnamefont {Vanderbilt}},\ }\href
  {\doibase 10.1038/s41467-021-24276-5} {\bibfield  {journal} {\bibinfo
  {journal} {Nature Communications}\ }\textbf {\bibinfo {volume} {12}},\
  \bibinfo {pages} {3998} (\bibinfo {year} {2021})}\BibitemShut {NoStop}%
\bibitem [{\citenamefont {Morano}\ \emph {et~al.}(2023)\citenamefont {Morano},
  \citenamefont {Gaudet}, \citenamefont {Varnava}, \citenamefont {Berry},
  \citenamefont {Halloran}, \citenamefont {Lygouras}, \citenamefont {Wang},
  \citenamefont {Hoffman}, \citenamefont {Xu}, \citenamefont {Lynn} \emph
  {et~al.}}]{Morano2023}%
  \BibitemOpen
  \bibfield  {author} {\bibinfo {author} {\bibfnamefont {V.~C.}\ \bibnamefont
  {Morano}}, \bibinfo {author} {\bibfnamefont {J.}~\bibnamefont {Gaudet}},
  \bibinfo {author} {\bibfnamefont {N.}~\bibnamefont {Varnava}}, \bibinfo
  {author} {\bibfnamefont {T.}~\bibnamefont {Berry}}, \bibinfo {author}
  {\bibfnamefont {T.}~\bibnamefont {Halloran}}, \bibinfo {author}
  {\bibfnamefont {C.~J.}\ \bibnamefont {Lygouras}}, \bibinfo {author}
  {\bibfnamefont {X.}~\bibnamefont {Wang}}, \bibinfo {author} {\bibfnamefont
  {C.~M.}\ \bibnamefont {Hoffman}}, \bibinfo {author} {\bibfnamefont
  {G.}~\bibnamefont {Xu}}, \bibinfo {author} {\bibfnamefont {J.~W.}\
  \bibnamefont {Lynn}},  \emph {et~al.},\ }\href@noop {} {\bibfield  {journal}
  {\bibinfo  {journal} {arXiv preprint arXiv:2311.00622}\ } (\bibinfo {year}
  {2023})}\BibitemShut {NoStop}%
\bibitem [{\citenamefont {Fender}\ \emph {et~al.}(2021)\citenamefont {Fender},
  \citenamefont {Thomas}, \citenamefont {Ronning}, \citenamefont {Bauer},
  \citenamefont {Thompson},\ and\ \citenamefont {Rosa}}]{Fender2021}%
  \BibitemOpen
  \bibfield  {author} {\bibinfo {author} {\bibfnamefont {S.~S.}\ \bibnamefont
  {Fender}}, \bibinfo {author} {\bibfnamefont {S.~M.}\ \bibnamefont {Thomas}},
  \bibinfo {author} {\bibfnamefont {F.}~\bibnamefont {Ronning}}, \bibinfo
  {author} {\bibfnamefont {E.~D.}\ \bibnamefont {Bauer}}, \bibinfo {author}
  {\bibfnamefont {J.~D.}\ \bibnamefont {Thompson}}, \ and\ \bibinfo {author}
  {\bibfnamefont {P.~F.~S.}\ \bibnamefont {Rosa}},\ }\href {\doibase
  10.1103/PhysRevMaterials.5.074603} {\bibfield  {journal} {\bibinfo  {journal}
  {Phys. Rev. Materials}\ }\textbf {\bibinfo {volume} {5}},\ \bibinfo {pages}
  {074603} (\bibinfo {year} {2021})}\BibitemShut {NoStop}%
\bibitem [{\citenamefont {Chapon}\ \emph {et~al.}(2011)\citenamefont {Chapon},
  \citenamefont {Manuel}, \citenamefont {Radaelli}, \citenamefont {Benson},
  \citenamefont {Perrott}, \citenamefont {Ansell}, \citenamefont {Rhodes},
  \citenamefont {Raspino}, \citenamefont {Duxbury}, \citenamefont {Spill},\
  and\ \citenamefont {Norris}}]{Chapon2011}%
  \BibitemOpen
  \bibfield  {author} {\bibinfo {author} {\bibfnamefont {L.~C.}\ \bibnamefont
  {Chapon}}, \bibinfo {author} {\bibfnamefont {P.}~\bibnamefont {Manuel}},
  \bibinfo {author} {\bibfnamefont {P.~G.}\ \bibnamefont {Radaelli}}, \bibinfo
  {author} {\bibfnamefont {C.}~\bibnamefont {Benson}}, \bibinfo {author}
  {\bibfnamefont {L.}~\bibnamefont {Perrott}}, \bibinfo {author} {\bibfnamefont
  {S.}~\bibnamefont {Ansell}}, \bibinfo {author} {\bibfnamefont {N.~J.}\
  \bibnamefont {Rhodes}}, \bibinfo {author} {\bibfnamefont {D.}~\bibnamefont
  {Raspino}}, \bibinfo {author} {\bibfnamefont {D.}~\bibnamefont {Duxbury}},
  \bibinfo {author} {\bibfnamefont {E.}~\bibnamefont {Spill}}, \ and\ \bibinfo
  {author} {\bibfnamefont {J.}~\bibnamefont {Norris}},\ }\href {\doibase
  10.1080/10448632.2011.569650} {\bibfield  {journal} {\bibinfo  {journal}
  {Neutron News}\ }\textbf {\bibinfo {volume} {22}},\ \bibinfo {pages} {22}
  (\bibinfo {year} {2011})},\ \Eprint
  {http://arxiv.org/abs/https://doi.org/10.1080/10448632.2011.569650}
  {https://doi.org/10.1080/10448632.2011.569650} \BibitemShut {NoStop}%
\bibitem [{\citenamefont {Arnold}\ \emph {et~al.}(2014)\citenamefont {Arnold},
  \citenamefont {Bilheux}, \citenamefont {Borreguero}, \citenamefont {Buts},
  \citenamefont {Campbell}, \citenamefont {Chapon}, \citenamefont {Doucet},
  \citenamefont {Draper}, \citenamefont {{Ferraz Leal}}, \citenamefont {Gigg},
  \citenamefont {Lynch}, \citenamefont {Markvardsen}, \citenamefont
  {Mikkelson}, \citenamefont {Mikkelson}, \citenamefont {Miller}, \citenamefont
  {Palmen}, \citenamefont {Parker}, \citenamefont {Passos}, \citenamefont
  {Perring}, \citenamefont {Peterson}, \citenamefont {Ren}, \citenamefont
  {Reuter}, \citenamefont {Savici}, \citenamefont {Taylor}, \citenamefont
  {Taylor}, \citenamefont {Tolchenov}, \citenamefont {Zhou},\ and\
  \citenamefont {Zikovsky}}]{Arnold2014}%
  \BibitemOpen
  \bibfield  {author} {\bibinfo {author} {\bibfnamefont {O.}~\bibnamefont
  {Arnold}}, \bibinfo {author} {\bibfnamefont {J.}~\bibnamefont {Bilheux}},
  \bibinfo {author} {\bibfnamefont {J.}~\bibnamefont {Borreguero}}, \bibinfo
  {author} {\bibfnamefont {A.}~\bibnamefont {Buts}}, \bibinfo {author}
  {\bibfnamefont {S.}~\bibnamefont {Campbell}}, \bibinfo {author}
  {\bibfnamefont {L.}~\bibnamefont {Chapon}}, \bibinfo {author} {\bibfnamefont
  {M.}~\bibnamefont {Doucet}}, \bibinfo {author} {\bibfnamefont
  {N.}~\bibnamefont {Draper}}, \bibinfo {author} {\bibfnamefont
  {R.}~\bibnamefont {{Ferraz Leal}}}, \bibinfo {author} {\bibfnamefont
  {M.}~\bibnamefont {Gigg}}, \bibinfo {author} {\bibfnamefont {V.}~\bibnamefont
  {Lynch}}, \bibinfo {author} {\bibfnamefont {A.}~\bibnamefont {Markvardsen}},
  \bibinfo {author} {\bibfnamefont {D.}~\bibnamefont {Mikkelson}}, \bibinfo
  {author} {\bibfnamefont {R.}~\bibnamefont {Mikkelson}}, \bibinfo {author}
  {\bibfnamefont {R.}~\bibnamefont {Miller}}, \bibinfo {author} {\bibfnamefont
  {K.}~\bibnamefont {Palmen}}, \bibinfo {author} {\bibfnamefont
  {P.}~\bibnamefont {Parker}}, \bibinfo {author} {\bibfnamefont
  {G.}~\bibnamefont {Passos}}, \bibinfo {author} {\bibfnamefont
  {T.}~\bibnamefont {Perring}}, \bibinfo {author} {\bibfnamefont
  {P.}~\bibnamefont {Peterson}}, \bibinfo {author} {\bibfnamefont
  {S.}~\bibnamefont {Ren}}, \bibinfo {author} {\bibfnamefont {M.}~\bibnamefont
  {Reuter}}, \bibinfo {author} {\bibfnamefont {A.}~\bibnamefont {Savici}},
  \bibinfo {author} {\bibfnamefont {J.}~\bibnamefont {Taylor}}, \bibinfo
  {author} {\bibfnamefont {R.}~\bibnamefont {Taylor}}, \bibinfo {author}
  {\bibfnamefont {R.}~\bibnamefont {Tolchenov}}, \bibinfo {author}
  {\bibfnamefont {W.}~\bibnamefont {Zhou}}, \ and\ \bibinfo {author}
  {\bibfnamefont {J.}~\bibnamefont {Zikovsky}},\ }\href {\doibase
  https://doi.org/10.1016/j.nima.2014.07.029} {\bibfield  {journal} {\bibinfo
  {journal} {Nuclear Instruments and Methods in Physics Research Section A:
  Accelerators, Spectrometers, Detectors and Associated Equipment}\ }\textbf
  {\bibinfo {volume} {764}},\ \bibinfo {pages} {156 } (\bibinfo {year}
  {2014})}\BibitemShut {NoStop}%
\bibitem [{\citenamefont {Rodr{\'\i}guez-Carvajal}(1993)}]{Rodriguez1993}%
  \BibitemOpen
  \bibfield  {author} {\bibinfo {author} {\bibfnamefont {J.}~\bibnamefont
  {Rodr{\'\i}guez-Carvajal}},\ }\href {\doibase
  https://doi.org/10.1016/0921-4526(93)90108-I} {\bibfield  {journal} {\bibinfo
   {journal} {Physica B: Condensed Matter}\ }\textbf {\bibinfo {volume}
  {192}},\ \bibinfo {pages} {55 } (\bibinfo {year} {1993})}\BibitemShut
  {NoStop}%
\bibitem [{\citenamefont {Collins}\ \emph {et~al.}(2010)\citenamefont
  {Collins}, \citenamefont {Bombardi}, \citenamefont {Marshall}, \citenamefont
  {Williams}, \citenamefont {Barlow}, \citenamefont {Day}, \citenamefont
  {Pearson}, \citenamefont {Woolliscroft}, \citenamefont {Walton},
  \citenamefont {Beutier},\ and\ \citenamefont {Nisbet}}]{Collins2010}%
  \BibitemOpen
  \bibfield  {author} {\bibinfo {author} {\bibfnamefont {S.~P.}\ \bibnamefont
  {Collins}}, \bibinfo {author} {\bibfnamefont {A.}~\bibnamefont {Bombardi}},
  \bibinfo {author} {\bibfnamefont {A.~R.}\ \bibnamefont {Marshall}}, \bibinfo
  {author} {\bibfnamefont {J.~H.}\ \bibnamefont {Williams}}, \bibinfo {author}
  {\bibfnamefont {G.}~\bibnamefont {Barlow}}, \bibinfo {author} {\bibfnamefont
  {A.~G.}\ \bibnamefont {Day}}, \bibinfo {author} {\bibfnamefont {M.~R.}\
  \bibnamefont {Pearson}}, \bibinfo {author} {\bibfnamefont {R.~J.}\
  \bibnamefont {Woolliscroft}}, \bibinfo {author} {\bibfnamefont {R.~D.}\
  \bibnamefont {Walton}}, \bibinfo {author} {\bibfnamefont {G.}~\bibnamefont
  {Beutier}}, \ and\ \bibinfo {author} {\bibfnamefont {G.}~\bibnamefont
  {Nisbet}},\ }\href {\doibase 10.1063/1.3463196} {\bibfield  {journal}
  {\bibinfo  {journal} {AIP Conference Proceedings}\ }\textbf {\bibinfo
  {volume} {1234}},\ \bibinfo {pages} {303} (\bibinfo {year} {2010})},\ \Eprint
  {http://arxiv.org/abs/https://pubs.aip.org/aip/acp/article-pdf/1234/1/303/11959497/303\_1\_online.pdf}
  {https://pubs.aip.org/aip/acp/article-pdf/1234/1/303/11959497/303\_1\_online.pdf}
  \BibitemShut {NoStop}%
\bibitem [{\citenamefont {Strempfer}\ \emph {et~al.}(2013)\citenamefont
  {Strempfer}, \citenamefont {Francoual}, \citenamefont {Reuther},
  \citenamefont {Shukla}, \citenamefont {Skaugen}, \citenamefont
  {Schulte-Schrepping}, \citenamefont {Kracht},\ and\ \citenamefont
  {Franz}}]{Strempfer2013}%
  \BibitemOpen
  \bibfield  {author} {\bibinfo {author} {\bibfnamefont {J.}~\bibnamefont
  {Strempfer}}, \bibinfo {author} {\bibfnamefont {S.}~\bibnamefont
  {Francoual}}, \bibinfo {author} {\bibfnamefont {D.}~\bibnamefont {Reuther}},
  \bibinfo {author} {\bibfnamefont {D.~K.}\ \bibnamefont {Shukla}}, \bibinfo
  {author} {\bibfnamefont {A.}~\bibnamefont {Skaugen}}, \bibinfo {author}
  {\bibfnamefont {H.}~\bibnamefont {Schulte-Schrepping}}, \bibinfo {author}
  {\bibfnamefont {T.}~\bibnamefont {Kracht}}, \ and\ \bibinfo {author}
  {\bibfnamefont {H.}~\bibnamefont {Franz}},\ }\href {\doibase
  10.1107/S0909049513009011} {\bibfield  {journal} {\bibinfo  {journal}
  {Journal of Synchrotron Radiation}\ }\textbf {\bibinfo {volume} {20}},\
  \bibinfo {pages} {541} (\bibinfo {year} {2013})}\BibitemShut {NoStop}%
\bibitem [{\citenamefont {Francoual}\ \emph {et~al.}(2013)\citenamefont
  {Francoual}, \citenamefont {Strempfer}, \citenamefont {Reuther},
  \citenamefont {Shukla},\ and\ \citenamefont {Skaugen}}]{Francoual2013}%
  \BibitemOpen
  \bibfield  {author} {\bibinfo {author} {\bibfnamefont {S.}~\bibnamefont
  {Francoual}}, \bibinfo {author} {\bibfnamefont {J.}~\bibnamefont
  {Strempfer}}, \bibinfo {author} {\bibfnamefont {D.}~\bibnamefont {Reuther}},
  \bibinfo {author} {\bibfnamefont {D.~K.}\ \bibnamefont {Shukla}}, \ and\
  \bibinfo {author} {\bibfnamefont {A.}~\bibnamefont {Skaugen}},\ }\href
  {\doibase 10.1088/1742-6596/425/13/132010} {\bibfield  {journal} {\bibinfo
  {journal} {Journal of Physics: Conference Series}\ }\textbf {\bibinfo
  {volume} {425}},\ \bibinfo {pages} {132010} (\bibinfo {year}
  {2013})}\BibitemShut {NoStop}%
\end{thebibliography}%



\section*{Supplementary material}

# Supplemental Material:

## Unusual magnetism of the axion-insulator candidate $Eu_5In_2Sb_6$

**Contents**

# 1. Representational analysis of magnetism in Eu₅In₂Sb₆

The results of the representational analysis are summarised in Fig. S1. The structural unit cell of Eu₅In₂Sb₆ contains ten $Eu^{2+}$ ions across two 4-fold and one 2-fold Wyckoff sites (Eu1, Eu2, Eu3), as illustrated in the figure. All ions lie in the $a$–$b$ plane ($z = 0$). The analyses of the observed magnetic propagation vectors $\mathbf{q}_m^\Gamma$ and $\mathbf{q}_m^Z$ in the space group $Pbam$ each yield eight irreducible representations (irreps). Since the magnetic basis vectors (which define the phase relations between the cartesian components of magnetic moments at each side) are the same for both the $\Gamma$ and $Z$ irreps, we list them together as $(\Gamma|Z)^{+/-}_{1,2,3,4}$. The choice of irrep and the spin-orientations $\phi = (\varphi_1, \varphi_2, \varphi_3)$ at the three Eu ions (Eu1_1, Eu2_1, Eu3_1) thus fully defines the magnetic order. The $\Gamma$ and $(Z)$ versions of this structure then differ in the (anti) parallel stacking of this coplanar motif along the $c$-axis.

The following considerations apply:

- It can be clearly inferred from bulk magnetometry that Eu₅In₂Sb₆ is an easy-plane system (neutron and x-ray scattering likewise confirm that the magnetic moments have no components along the $c$-axis). This rules out $(\Gamma|Z)_1^+$, $(\Gamma|Z)_2^+$, $(\Gamma|Z)_3^-$ and $(\Gamma|Z)_4^-$, for which all moments must be (anti-)aligned with the $c$-axis.

- Similarly, bulk magnetometry and X-ray absorption spectra (see Supplementary Note 8) confirm that all magnetic sites must feature an ordered $Eu^{2+}$ magnetic moment. This rules out irreps $(\Gamma|Z)^+_{1,2,3,4}$, which require the Eu2 (Wyckoff $2a$) site to be non-magnetic.

The directions of magnetic structure factors determined in neutron and X-ray scattering experiments (see sections below) allow to unequivocally rule out $(\Gamma|Z)_4^+$ (see below), which leaves $(\Gamma|Z)_3^+$ as the only choice.

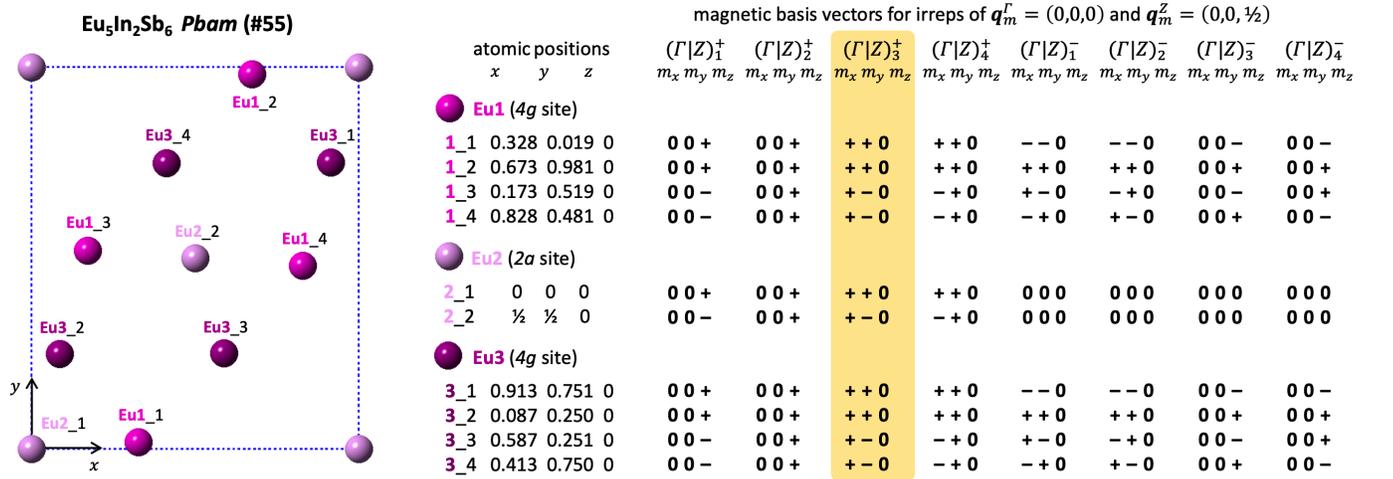

**Eu₅In₂Sb₆ _Pbam_ (#55)**

magnetic basis vectors for irreps of $\boldsymbol{q}_m^\Gamma = (0,0,0)$ and $\boldsymbol{q}_m^Z = (0,0,\frac{1}{2})$

| atomic positions | | | $(\Gamma\|Z)_1^+$ | $(\Gamma\|Z)_2^+$ | $(\Gamma\|Z)_3^+$ | $(\Gamma\|Z)_4^+$ | $(\Gamma\|Z)_1^-$ | $(\Gamma\|Z)_2^-$ | $(\Gamma\|Z)_3^-$ | $(\Gamma\|Z)_4^-$ |
| --- | --- | --- | --- | --- | --- | --- | --- | --- | --- | --- |
| | $x$ | $y$ | $m_x\,m_y\,m_z$ | $m_x\,m_y\,m_z$ | $m_x\,m_y\,m_z$ | $m_x\,m_y\,m_z$ | $m_x\,m_y\,m_z$ | $m_x\,m_y\,m_z$ | $m_x\,m_y\,m_z$ | $m_x\,m_y\,m_z$ |
| **Eu1 (_4g_ site)** | | | | | | | | | | |
| 1_1 | 0.328 | 0.019 | 00+ | 00+ | ++0 | ++0 | --0 | --0 | 00- | 00- |
| 1_2 | 0.673 | 0.981 | 00+ | 00+ | ++0 | ++0 | ++0 | ++0 | 00+ | 00+ |
| 1_3 | 0.173 | 0.519 | 00- | 00+ | +-0 | -+0 | +-0 | -+0 | 00- | 00+ |
| 1_4 | 0.828 | 0.481 | 00- | 00+ | +-0 | -+0 | -+0 | +-0 | 00+ | 00- |
| **Eu2 (_2a_ site)** | | | | | | | | | | |
| 2_1 | 0 | 0 | 00+ | 00+ | ++0 | ++0 | 000 | 000 | 000 | 000 |
| 2_2 | ½ | ½ | 00- | 00+ | +-0 | -+0 | 000 | 000 | 000 | 000 |
| **Eu3 (_4g_ site)** | | | | | | | | | | |
| 3_1 | 0.913 | 0.751 | 00+ | 00+ | ++0 | ++0 | --0 | --0 | 00- | 00- |
| 3_2 | 0.087 | 0.250 | 00+ | 00+ | ++0 | ++0 | ++0 | ++0 | 00+ | 00+ |
| 3_3 | 0.587 | 0.251 | 00- | 00+ | +-0 | -+0 | +-0 | -+0 | 00- | 00+ |
| 3_4 | 0.413 | 0.750 | 00- | 00+ | +-0 | -+0 | -+0 | +-0 | 00+ | 00- |

Fig. S1 Overview of the results of representational analysis of the magnetic propagation vectors $\mathbf{q}_m^\Gamma$ and $\mathbf{q}_m^Z$ in the space group $Pbam$ of Eu₅In₂Sb₆. The position and conventional numbering of the ten $Eu^{2+}$ magnetic ions in the structural unit cell are shown (all positions at $z = 0$). The table lists the phase relations of the cartesian components of magnetic moments (_magnetic basis vectors_) that are defined by each of the eight irreps. These basis vectors are the same for $\mathbf{q}_m^\Gamma$ and $\mathbf{q}_m^Z$, i.e. the respective magnetic structures vary only in terms of the stacking of the "per-plane" atomic arrangement along the $c$-axis.

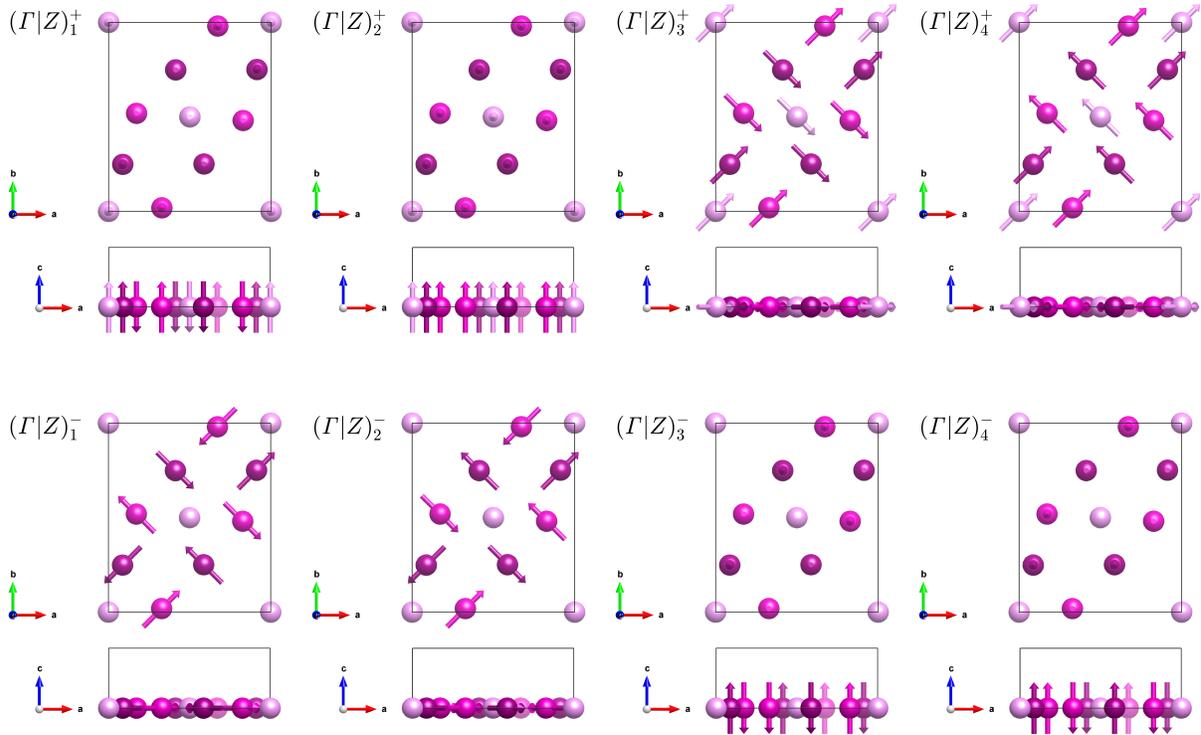

Fig. S2 Illustrations of the basis vectors corresponding to the eight irreducible representation of $\mathbf{q}_m^{\Gamma}$ and $\mathbf{q}_m^{Z}$ in the space group $Pbam$, as stated in the table in Fig. S1. For each basis vector, cartesian components of equal magnitude are drawn (where they do not vanish), i.e. $\mathbf{m}_{\text{Eu1,Eu2,Eu3}} = (m_x, m_y, m_z) \propto (1, 1, 1)$. Note that there are no symmetry constraints on the relative orientation of $\mathbf{m}_{\text{Eu1}}$, $\mathbf{m}_{\text{Eu2}}$ and $\mathbf{m}_{\text{Eu3}}$.

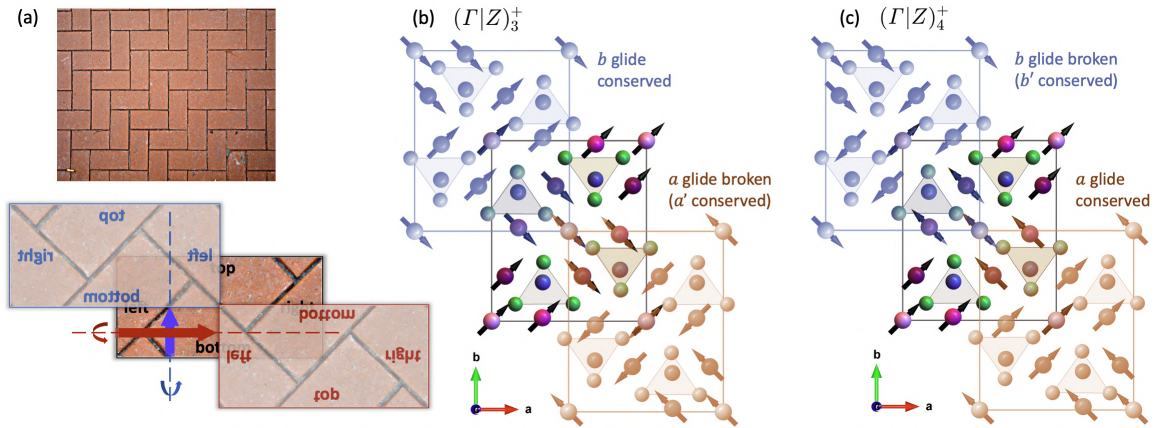

Fig. S3 (a) The arrangement of the Eu ions corresponds to the wallpaper group $pgg$, known as herringbone tiling. The position and effect of the perpendicular glide symmetries are illustrated on a "unit cell" of this pattern. (b) Corresponding illustration of the $(\Gamma|Z)_3^+$ arrangement as shown in Fig. S2: The magnetic order breaks the $a$-glide and conserves the $b$-glide. These types of structures are realized in $\text{Eu}_5\text{In}_2\text{Sb}_6$. (c) Conversely, $(\Gamma|Z)_4^+$ (ruled out by experiment) conserves the $a$-glide.

## 2. Magnetic structure factor vector

Regardless of whether the $\mathbf{q}_m^\Gamma$ or the $\mathbf{q}_m^Z$ component is being considered, the structure factor vector $\mathbf{M}$ is proportional to

$$\mathbf{M} \propto \sum_{i=1}^{10} \mathbf{m}_i \, e^{i\,2\pi \mathbf{Q}\cdot\mathbf{r}} \tag{1}$$

Where the sum runs over the ten ionic positions in the structural unit cell, as listed in Fig. S1 (the relation holds for both $\Gamma$ and $Z$ irreps). Evaluation Eq. 1 with the magnetic phase relations defined by the irrep candidates $(\Gamma|Z)_3^+$, and $(\Gamma|Z)_4^+$ yields

$$\mathbf{M} \propto \begin{cases} \begin{pmatrix} +2\cos(\varphi_1)\,c_{x1}\,c_{y1} & +\cos(\varphi_2) & +2\cos(\varphi_3)\,c_{x3}\,c_{y3} \\ -2\sin(\varphi_1)\,s_{x1}\,s_{y1} & & -2\sin(\varphi_3)\,s_{x3}\,s_{y3} \\ & 0 & \end{pmatrix} & \begin{array}{l} \text{for } (\Gamma|Z)_3^+,\ \text{if} \\ (h+k)\ \text{even} \end{array} \ \ \text{or} \ \ \begin{array}{l} \text{for } (\Gamma|Z)_4^+,\ \text{if} \\ (h+k)\ \text{odd} \end{array} \\[2em] \begin{pmatrix} -2\cos(\varphi_1)\,s_{x1}\,s_{y1} & & -2\cos(\varphi_3)\,s_{x3}\,s_{y3} \\ +2\sin(\varphi_1)\,c_{x1}\,c_{y1} & +\sin(\varphi_2) & +2\sin(\varphi_3)\,c_{x3}\,c_{y3} \\ & 0 & \end{pmatrix} & \begin{array}{l} \text{for } (\Gamma|Z)_3^+,\ \text{if} \\ (h+k)\ \text{odd} \end{array} \ \ \text{or} \ \ \begin{array}{l} \text{for } (\Gamma|Z)_4^+,\ \text{if} \\ (h+k)\ \text{even} \end{array} \end{cases} \tag{2}$$

Here we have used the following abbreviations:

$$s_{xi} = \sin(2\pi\,h\,x_i) \text{ and } c_{xi} = \cos(2\pi\,h\,x_i)\ , \qquad \text{with } x_1 = 0.328 \text{ and } x_3 = 0.913$$
$$s_{yi} = \sin(2\pi\,k\,y_i) \text{ and } c_{yi} = \cos(2\pi\,k\,y_i)\ , \qquad \text{with } y_1 = 0.019 \text{ and } y_3 = 0.751$$

$\varphi_{1,2,3}$ refer to the angle between the $a$ axis and the (in-plane) magnetic moments at the positions Eu1-1, Eu2-1 and Eu3-1. For magnetic Bragg reflections $(h,k,l)$ where one of the in-plane Bragg indices ($h$ or $k$) vanishes, Eq. 2 corresponds to the rule summarized in Table I.

| $h$ or $k$ | $(\Gamma|Z)_3^+$ | $(\Gamma|Z)_4^+$ |
|---|---|---|
| even | $\mathbf{M} \parallel \mathbf{a}$ | $\mathrm{M} \parallel \mathrm{b}$ |
| odd | $\mathbf{M} \parallel \mathbf{b}$ | $\mathrm{M} \parallel \mathrm{a}$ |

TABLE I: Directions of the magnetic structure factor vector $\mathbf{M}$, as given in Eq. 2, for cases where one of the in-plane Bragg indices ($h$ or $k$) vanishes. I.e. $(h,0,l)$ and $(0,k,l)$ peaks in the $\Gamma$ phase or $(h,0,l+\text{½})$ and $(0,k,l+\text{½})$ peaks in the $Z$ phase. The two candidate irreps $(\Gamma|Z)_3^+$ and $(\Gamma|Z)_4^+$ can thus be clearly distinguished. The experimental data clearly rules out $(\Gamma|Z)_4^+$.

The observations from both x-ray and neutron scattering unequivocally rule out $(\Gamma|Z)_4^+$. For a complete knowledge of the magnetic order in $(\Gamma|Z)_3^+$, the three degrees of freedom $\phi_{1,2,3}$ remain to be determined.

### 3. Neutron powder diffraction (WISH, ISIS)

Time-of-flight neutron powder diffraction (NPD) histograms of $Eu_5In_2Sb_6$ were obtained in the paramagnetic state (25 K), in the $\Gamma$ phase (10 K), and in the low-temperature $\Gamma/Z$ mixed state (1.5 K). These measurements are extremely challenging due to the overwhelming neutron absorption cross section of Eu. An analytical absorption correction [1] was included as a free parameter in the refinements but does not necessarily provide an accurate model. The results therefore carry an unknown systematic uncertainty. Nevertheless, all refinements of spin configurations reproducibly converged at global minima of the least squares.

**Intermediate-temperature phase ($\Gamma$ component only)**

The bulk magnetization of an $Eu_5In_2Sb_6$ single crystal measured in a magnetic field $\mathbf{H} \parallel \mathbf{a}$ at 10 K showed a hysteresis with a remanent magnetization of only $m_{net} \approx 0.11\,\mu_B$ [2]. This indicates that the spin orientation must be $\varphi_i \approx \pm 90°$, i.e. largely along the $b$-axis. There are four distinct configurations that can be constructed from $\varphi_{1,2,3} = \pm 90°$, A:"$+++$", B:"$+--$", C:"$+-+$", D:"$++-$". Out of these, all except B clearly contradict the magnetic intensities observed in NPD.

Fig. S4 illustrates two spin arrangements that both provide an acceptable model of the neutron data of the $\Gamma$ spin arrangement at 10 K. A free refinement of the moment directions yields significant canting away from the collinear structure B, $\phi = (+90°, -90°, -90°) + (45°, 21°, 6°)$, and least squares $\chi^2 = 1.22$. However, restricting the spin orientations to the collinear solution B also provides an acceptable fit ($\chi^2 = 1.56$), but at different scale factor, ordered magnetic moment, and absorption correction factor. The two solutions shown in Fig. S4 thus give an estimate of the uncertainty due to the correlation of these parameters.

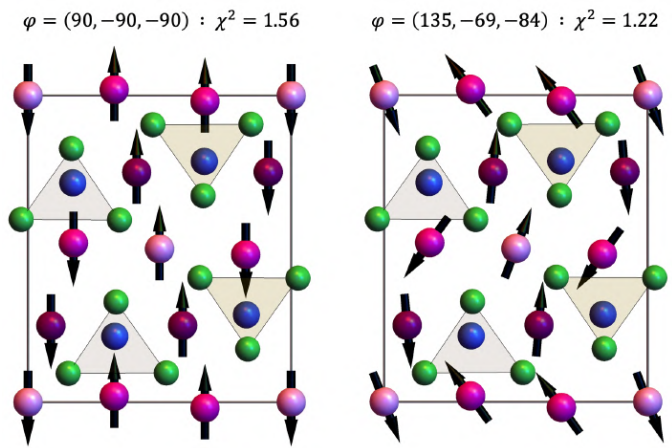

$\varphi = (90, -90, -90) : \chi^2 = 1.56$    $\varphi = (135, -69, -84) : \chi^2 = 1.22$

Fig. S4. $\Gamma$-type spin arrangements compatible with the neutron powder diffraction data at 10 K.

**Low-temperature $\Gamma/Z$ mixed phase**

Only components of the magnetic moments perpendicular to the momentum transfer contribute to magnetic neutron scattering intensities, and magnetic intensity at small $d$ spacing is suppressed by the $Eu^{2+}$ magnetic form factor, as indicated in Fig. S5(a). In our 1.5 K dataset, shown in Fig. S5(b), significant magnetic scattering from the $Z$ phase is observed at three positions:

1. $\mathbf{Q} = (0, 0, 1/2)$, at $d = 9.25\,\text{Å}$

2. $\mathbf{Q} = (0, 2, 1/2)$, at $d = 5.73\,\text{Å}$  [coincident with $(2, 1, 0)$ magnetic intensity of the $\Gamma$ phase]

3. $\mathbf{Q} = (0, 4, 1/2)$, at $d = 3.39\,\text{Å}$

The common characteristic of these peaks is that $h = 0$ and $k$ is even, hence $\mathbf{M} \parallel \hat{\mathbf{a}}$ (see Supplementary Note 2), and hence in each case $\mathbf{Q} \perp \mathbf{M}$, i.e. these are positions that are sensitive to magnetic components parallel to the $a$ axis. Crucially, we do *not* observe magnetic intensity at $\mathbf{Q} = (1, 0, 1/2)$, where $\mathbf{Q} \perp \mathbf{M} \parallel \hat{\mathbf{b}}$, which lies at a $d = 7.44\,\text{Å}$, i.e. favored by the form factor and sensitive to tilts of the $Z$ component away from the $a$ axis. The 1.5 K data can be modelled equally well, assuming either magnetic phase separation or a double-$q$ magnetic order (see below).

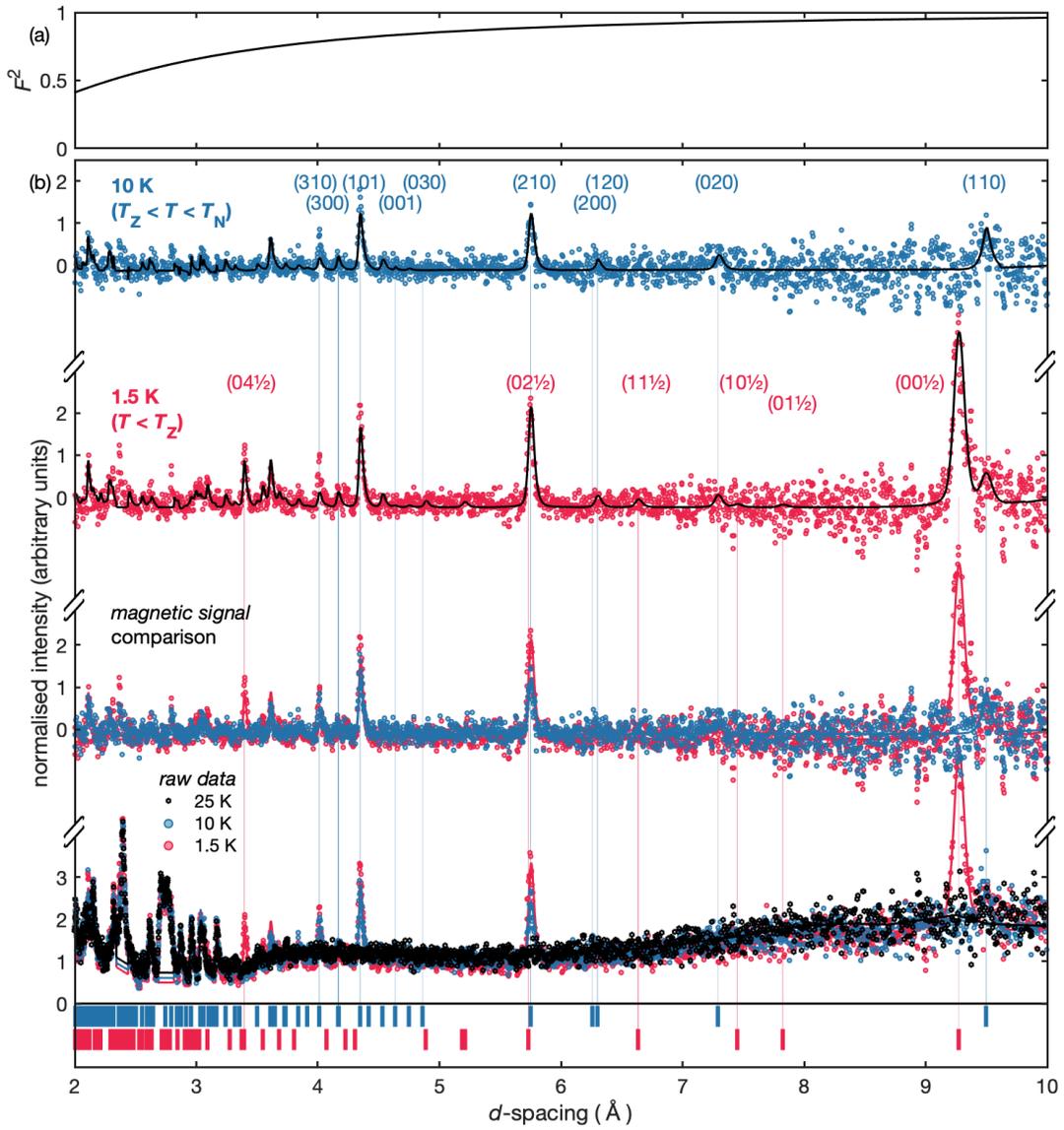

Fig. S5. Time-of-flight neutron powder diffraction of $Eu_5In_2Sb_6$. (a) Square of the $Eu^{2+}$ neutron magnetic form factor. (b) Neutron powder histograms recorded in the paramagnetic state (25 K), in the magnetic $\Gamma$ phase (10 K) and in the mixed $\Gamma/Z$ state (1.5 K). The raw data (bottom) is shown along with the difference patterns (10 K, 1.5 K) obtained by subtracting the nuclear scattering (25 K). The solid lines indicate best fits of the respective magnetic structure models.

## 4. Comparison of phase separation and double-$q$ models

Magnetic scattering reveals that the low-temperature ($T < T_Z = 7.5\,\text{K}$) state of $\text{Eu}_5\text{In}_2\text{Sb}_6$ features both $\Gamma$ and $Z$ Fourier components, associated with the propagation vectors $q_\Gamma = (0,0,0)$ and $q_Z = (0,0,1/2)$. There are two possible interpretations:

- **Phase separation.** A straightforward interpretation is the magnetic phase separation into single-$q$ magnetic regimes. Given that magnetometry indicates an ordered moment of $\approx 8\,\mu_\text{B}$ at each site [2], the degrees of freedom of this magnetic state are one $\Gamma/Z$ volume ratio, and three in-plane spin orientations for each phase, which we define by angles $\phi_{1,2,3}^\Gamma$ and $\phi_{1,2,3}^Z$ w.r.t. the $a$-axis.

- **Double-$q$ order.** Since the $\Gamma$ and $Z$ magnetic orders are defined by the same in-plane magnetic basis vectors (see Supplementary Note 2), the magnitude of the ordered magnetic moment can only be conserved in the special case that the $\Gamma$ and $Z$ Fourier components of each Wyckoff site are perpendicular (otherwise the magnitude of the ordered moment would vary from layer to layer). The degrees of freedom of this model are one spin-orientation and one $\Gamma/Z$ ratio $\eta_{1,2,3}$ per Wyckoff site, as defined below.

While the number of degrees of freedom of the two models is similar, they imply different constraints: For instance, in the phase separation model, the spin orientations $\phi_{1,2,3}^\Gamma$ and $\phi_{1,2,3}^Z$ are independent, but the magnitude of the associated components must be the same at each site. Conversely, in the double-$q$ model, the $\Gamma$ and $Z$ components are not independent (they must be perpendicular), but their relative size may vary across sites (e.g. one site may be purely $\Gamma$, another purely $Z$). In a diffraction experiment, the two models are, therefore, not necessarily indistinguishable.

Fig. S6 illustrates best fits to the $1.5\,\text{K}$ time-of-flight neutron powder diffraction data (collected on banks 2 and 9 of the WISH instrument). Magnetic phase separation, allowing tilts away from a collinear configuration, yields the structures illustrated in Fig. S6(d):

$$66\,\%\ \Gamma\ \text{phase},\qquad \phi_{1,2,3}^\Gamma = \{89(4)^\circ, 241(2)^\circ, 262(3)^\circ\}$$
$$34\,\%\ Z\ \text{phase},\qquad \phi_{1,2,3}^Z = \{2(11)^\circ, -22(6)^\circ, 20(4)^\circ\}$$

Note that the uncertainties quoted here represent the standard deviations of the Rietveld refinement (statistical uncertainty). As discussed in Supplementary Note 3, the phenomenological absorption correction may not adequately account for the strong neutron absorption of Eu. Comparison with the fit to a collinear configuration, see Fig. S6(a,b), illustrates the resulting systematic uncertainty.

For the double-$q$ model, we fixed $\phi_{1,2,3}^Z = \phi_{1,2,3}^\Gamma + 90^\circ$ and implemented the (non-linear) constraints on the magnitude of the magnetic components described above:

$$|M_{\text{Eu1,Eu2,Eu3}}^\Gamma| = M\,\sin\!\left(\frac{\pi}{2}\,\eta_{1,2,3}\right),$$
$$|M_{\text{Eu1,Eu2,Eu3}}^Z| = M\,\cos\!\left(\frac{\pi}{2}\,\eta_{1,2,3}\right),$$

where the factors $\eta_{1,2,3}$ describe the balance of $\Gamma$ and $Z$ character at each Wyckoff site and each site carries the same ordered moment, $\sqrt{(M_{\text{Eu1,Eu2,Eu3}}^\Gamma)^2 + (M_{\text{Eu1,Eu2,Eu3}}^Z)^2} = M$. Independently of the starting parameters $\eta_{1,2,3}$ and $\phi_{1,2,3}^\Gamma$, an optimization of these six degrees of freedom reproducibly converged at the unique solution that is illustrated in Fig. S6(f):

$$\eta_{1,2,3} = \{0.22, 1.31, 1.21\},\qquad \phi_{1,2,3}^\Gamma = \{94(5)^\circ, 248(2)^\circ, 268(3)^\circ\}$$

In summary, our neutron powder diffraction experiment provides a unique solution for either model of the low-temperature magnetic state of $\text{Eu}_5\text{In}_2\text{Sb}_6$. However, a distinction is not possible because the two solutions provide almost exactly the same fit to the data.

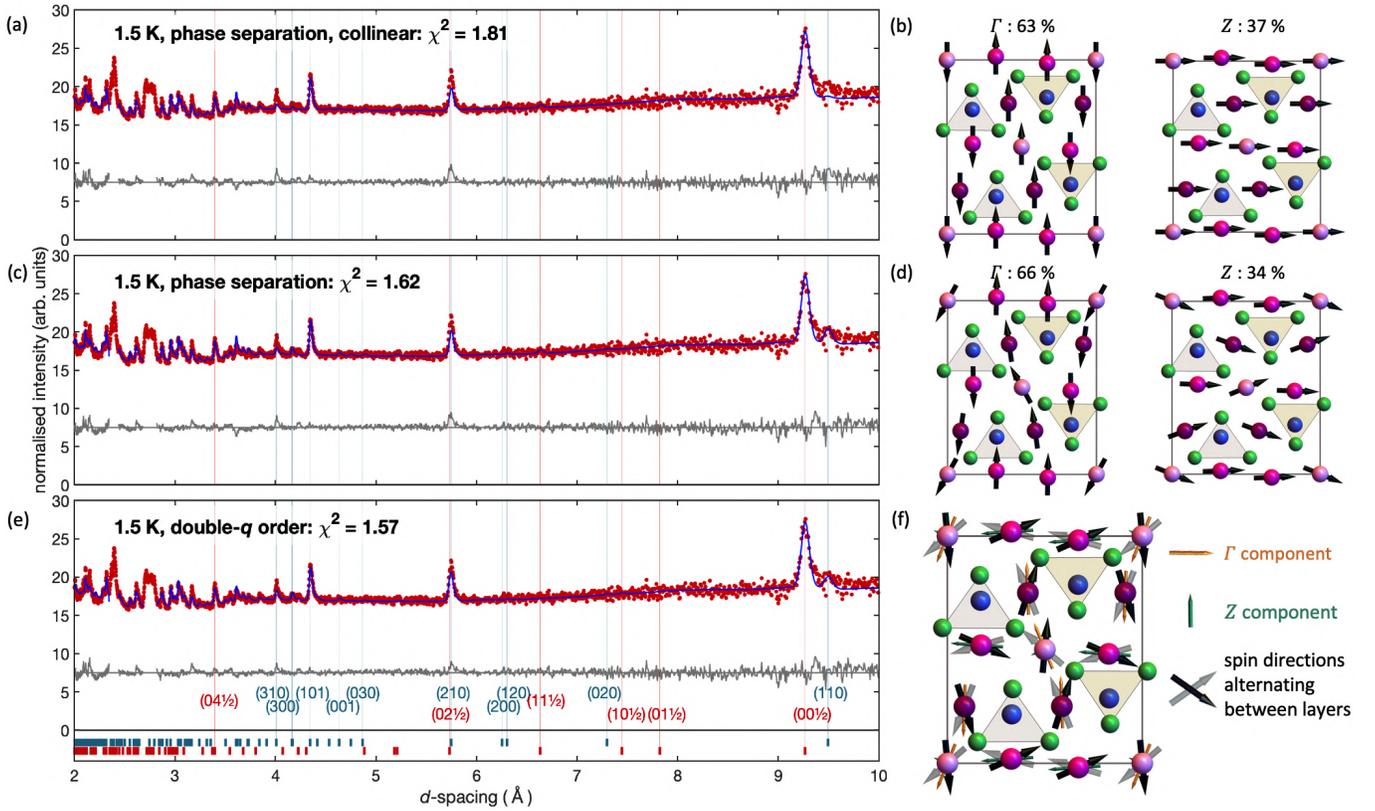

Fig. S6. Ambiguity of the 1.5 K neutron powder diffraction data with respect to two structural models. (a) Rietveld refinement, assuming magnetic phase separation into regimes with collinear single-$q$ order. The difference between data (red markers) and the best fit (blue line) is indicated as a grey line with an arbitrary offset. (b) The result corresponds to volume fractions of ca. two thirds $\Gamma$ phase ($q_\Gamma = (0,0,0)$), with the spin arrangements shown here. (c) Refinement of the same data, including symmetry-allowed tilts as free refinement parameters. (d) Illustration of the best fit in this model of phase separation and non-collinear spins. (e) Refinement of the same data, assuming double-$q$ magnetic order with certain physical constraints (see text). (f) The result corresponds to a single phase in which each of the three Eu sites features a different ratio of perpendicular $\Gamma$ and $Z$-type components, which add to the same ordered moment at each site.

## 5. Intermediate-temperature $\Gamma_3^+$ magnetic structure: Spin orientations from REXS FLPA

In REXS, magnetic scattering with a propagation vector $\mathbf{q}_m^\Gamma = (0,0,0)$ is only observable where charge (Thompson) scattering is extinct. The extinction conditions for Bragg peaks $hkl$ in space group $Pbam$ are

$$\left.\begin{array}{c} h00 \\ h0l \end{array}\right\} \text{extinct if } h \text{ odd} \qquad \left.\begin{array}{c} 0k0 \\ 0kl \end{array}\right\} \text{if } k \text{ odd} \tag{3}$$

Crucially, there are no "forbidden" peaks for which both $h \neq 0$ and $k \neq 0$. In other words, REXS can only observe the $\Gamma$ magnetic scattering at peaks where one of the in-plane indices is zero and the other is an odd integer. Given the $\Gamma_3^+$ rule derived above (Table I, Supplementary Note 2), this means that for any accessible peak, the magnetic structure factor is necessarily perfectly parallel to the $b$-axis. For instance, in our REXS study of the $\Gamma$ phase, we characterized the magnetic intensity at (3,0,0). According to Eq. 2, the magnetic structure factor is

$$\mathbf{M}_{(h,0,0)}^{\Gamma 3+} \propto \hat{\mathbf{b}} \; [2 \sin(\varphi_1) \cos(2\pi \, h \, x_1) + \sin(\varphi_2) + 2 \sin(\varphi_3) \cos(2\pi \, h \, x_3)]$$
$$\rightarrow \mathbf{M}_{(3,0,0)}^{\Gamma 3+} \propto \hat{\mathbf{b}} \; [1.990 \sin(\varphi_1) + \sin(\varphi_2) - 0.138 \sin(\varphi_3)] \quad,$$

where we have used the atomic $x$-coordinates of Eu1_1 ($x_1 = 0.328$) and Eu3_1 ($x_3 = 0.913$). This shows that the intensity of any accessible peak of the intermediate ($\Gamma$) phase is only a weighted measure of components of the $b$ axis components of the magnetic moments. For this state, it is therefore not possible to use the polarisation analysis of a single REXS peak to infer the relative weight of $b$ and $a$ axis components and thus to deduce the non-collinear in-plane spin arrangement, i.e. the in-plane angles $\vec{\phi} = (\varphi_1, \varphi_2, \varphi_3)$.

Fig. S7 shows REXS full linear polarisation analysis (FLPA) scans of the (3,0,0) magnetic intensity in the low-$T$ state at $5\,\mathrm{K}$ and in the intermediate state at $10\,\mathrm{K}$. For a series of orientations $\eta$ of the incident linear polarisation, the polarization angle $\eta'$ of the scattered beam is characterised by the Stokes parameters $P_1'$ and $P_2'$:

$$I(\eta') = I_0 + I_0 \; [P_1' \cos(2\eta') + P_2' \cos(2\eta')] \tag{4}$$

In these measurements, the crystal was mounted with the momentum transfer along the specular (1,0,0) direction, the $c$ axis perpendicular to the (vertical) scattering plane, and the $b$-axis parallel to $\hat{u}_2 = (\hat{k}_i \times \hat{k}_f)/\sin(2\theta)$. The lines drawn in Panels (b,d) indicate the characteristics of $P_1'$ and $P_2'$ expected for a magnetic structure factor $\hat{\mathbf{M}} \parallel \hat{b}$. Magnetic intensity is maximal in the $\pi\sigma'$ ($\eta = \pm 90°$, $\eta' = 0°/180°$) and $\sigma\pi'$ channels, where $\hat{M}$ is projected onto the vectors $\hat{k}_i$ and $-\hat{k}_f$, respectively. The resulting Stokes parameters $P' = (P_1', P_2')$ of the scattered beam are $(1,0)$ for $\pi\sigma'$ and $(-1,0)$ for $\sigma\pi'$.

The observations confirm that this phase is described by $\Gamma_3^+$. While its intensity decreases when cooling below $T < T_Z$, the FLPA characteristics show no qualitative changes. To infer more detailed information on the spin arrangement of this phase, we performed neutron powder diffraction, where several $\mathbf{q}_m^\Gamma = (0,0,0)$ magnetic intensities are measurable.

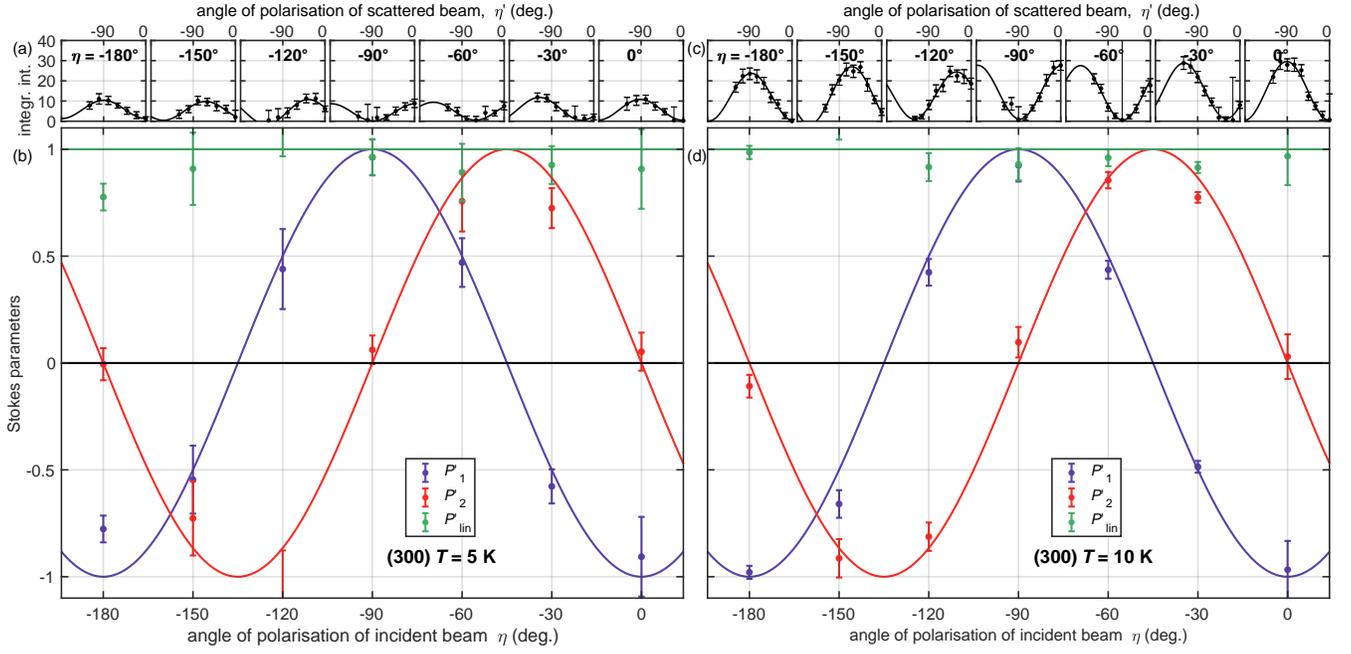

Fig. S7. Eu $L_3$ REXS full linear polarization analysis (FLPA) of magnetic intensity at the (300) peak of the $\Gamma_3^+$ magnetic structure. (a,b) Data recorded at 5 K (low-$T$ $(\Gamma|Z)_3^+$ mixed state), and (c,d) at 10 K (intermediate-$T$ $\Gamma_3^+$-only state). For a series of orientations $\eta$ of the linear polarisation of the incident beam, the upper panels (a,c) show the variation of the intensity measured for a certain orientation $\eta'$ of the polarization analyzer crystal. The lower panels (b,d) show the Stokes parameters $P'$ inferred from fits to this data. The fits confirm that the magnetic structure factor vector is constrained along the $b$ axis. While the intensities subside below $T_Z$ (due to the decreasing $\Gamma$-type magnetic component), the FLPA characteristics persist.

## 6. Low-temperature $Z_3^+$ magnetic component: Spin orientations from REXS FLPA

To determine the spin-orientations $\phi = (\varphi_1, \varphi_2, \varphi_3)$ in the low-temperature $(T < T_Z)$ $Z_3^+$ magnetic component, we performed a full linear polarisation analysis (FLPA) study of the resonant magnetic scattering at the Eu $L_3$ edge (instrument P09, DESY). The scattering geometry, results and analysis of this experiment are summarized in Fig. S8. The sample was mounted with the $b$ axis specular, which provides access to a number of magnetic Bragg peaks $(h, k, l + 0.5)$ with the momentum transfer close to the $(0,1,0)$ direction. We investigated five such peaks, each in two azimuthal orientations $\psi = 90°$ and $\psi = 45°$ (with the $c$-axis as azimuthal reference, see Fig. S8(a,c)).

To perform an FLPA, we selected angles $\eta$ of the incident linear polarisation and then recorded the diffracted intensity for a series of angles $\eta'$ of the polarisation analyzer. Panels (b,d) show the resulting Stokes parameters $\vec{P}' = (P_1', P_2')$ that describe the polarization of the scattered beam. $P_{\text{lin}}' = \sqrt{P_1'^2 + P_2'^2}$ is the degree of linear polarisation. The data is described by a global model of all recorded intensities with fit parameters $(\varphi_1, \varphi_2, \varphi_3)$ and an arbitrary scale parameter per peak. We find that this fit has four global minima corresponding to the four possible magnetic structures in which all moments are almost parallel to the $a$-axis. These four choices correspond to the four possible phase-relations between the three Wyckoff sites occupied by Eu ions (A:"$+++$", B:"$+--$", C:"$+-+$", D:"$++-$").

Fig. S8 (e) shows a least-squares ($\chi^2$) map of the $(-5.4°, \varphi_2, \varphi_3)$ plane, which contains configurations close to each of the four minima A–D. The resulting structures are not perfectly aligned with the $a$ axis, e.g. $\vec{\phi}_A = (-4.2, 11.5, 1.3)$ and $\vec{\phi}_B = (-5.4, 168.6, 178.4)$. This is illustrated in the detailed $\chi^2$-maps and confidence contours in Panels (d,e).

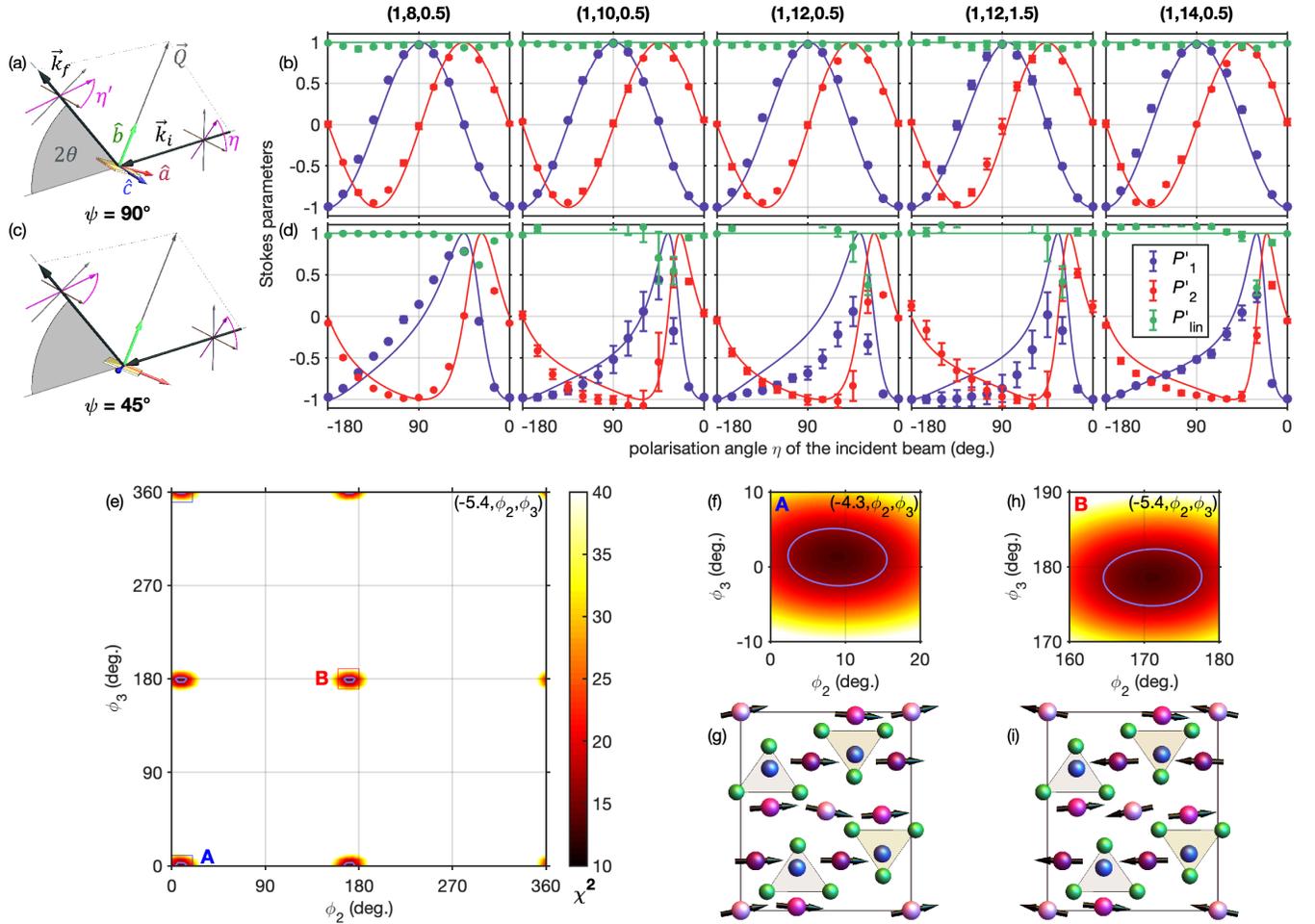

Fig. S8. Eu $L_3$ REXS full linear polarisation analysis of the $Z_3^+$ magnetic structure in Eu$_5$In$_2$Sb$_6$ at 4.6 K. The figure illustrates global fits of the polarisation characteristics of five $\langle 0, 0, 0.5 \rangle$-type magnetic Bragg peaks, each at two azimuthal orientations. (a,c) Illustration of the experimental geometry at P09 (DESY), with a vertical scattering plane and momentum transfers $\vec{Q}$ close to the $b$ axis, for $\psi = 90°$ and $\psi = 45°$. (b,d) Best fits of the Stokes parameters $P'$ of the scattered beam as a function of the polarization angle $\eta$ of the incident beam for $\psi = 90°$ (b) and $\psi = 45°$ (d). The lines show the calculated characteristics of the spin arrangement B with angles $\phi_B = (-5.4, 168.6, 178.4)$. (e) Color map of the $\phi = (-5.4°, \varphi_2, \varphi_3)$ plane illustrating the variation of the least squares $\chi^2$ of a global fit of all data. There are four global minima, corresponding to four arrangements with all moments almost parallel to the $a$ axis. (f,h) Detail views of the $\chi^2$ maps around the configurations A $\phi_A = (-4.2, 11.5, 1.3)$ and B . The blue contour indicates the confidence interval of one standard deviation. (g,i) Illustrations of the spin arrangements A and B in the $a$-$b$ plane.

### 7. Low-temperature $Z_3^+$ magnetic component: Spin orientations from REXS azimuthal scans

The spin alignment $\phi = (\varphi_1, \varphi_2, \varphi_3)$ in the $\Gamma/Z$ mixed state ($T < T_Z = 7.5\,\mathrm{K}$) can be determined with high accuracy using azimuthal scans of resonant elastic X-ray scattering. We performed REXS azimuthal ($\psi$) scans of several magnetic peaks close to the specular $(0, 1, 0)$ direction. These measurements were performed at $6\,\mathrm{K}$, i.e. just below $T_Z$. The intensity was measured in vertical scattering in the $\sigma\pi'$ polarisation channel. As illustrated in Fig. S9(a), in this configuration, the REXS intensity is proportional to the projection of the magnetic structure factor vector onto the direction of the scattered beam, $I \propto |\mathbf{M} \cdot \mathbf{k}_f|^2$. We define $(001)$ as the azimuthal reference so that $\psi = 0°$ when $(0,0,1)$ lies in the scattering plane and $\psi = 90°$ when $(1,0,0)$ lies in the scattering plane. In this configuration, the data can be read as follows:

- For peaks where $\mathbf{M} \parallel \hat{\mathbf{a}}$, the azimuthal scan with be a $\sin^2(\psi)$ function, i.e. it will vanish at $\psi = 0$ and will have equal maxima at $\psi \pm 90°$.

- For peaks where $\mathbf{M} \parallel \hat{\mathbf{b}}$, the azimuthal scan will show constant intensity.

- If $\mathbf{M}$ has non-vanishing components $M_x$ and $M_y$, the two maxima $\psi \approx \pm 90°$ will have different intensities, and the zero-point will shift from $\psi = 0$ towards the smaller maximum.

In Fig. S9, we summarize our analysis of the REXS $\sigma\pi'$ azimuthal scans of $(h, 14, \pm 0.5)$ magnetic peaks, for $h = -1, 0, +1$. For each peak, Panel (a) shows the magnetic structure factor vector and a sketch of the scattering geometry. The corresponding data and fits are given in Panel (b). For $h = 0$, the magnetic structure factor must lie along the $a$ axis (regardless of the spin orientations $\phi$) and hence the azimuthal scan shows $\sin^2(\psi)$ characteristics. For the peaks with $h \pm 1$, $\mathbf{M}$ has a non-vanishing component $\mathbf{M}_y$ and opposite components $\mp\mathbf{M}_x$. The projections of $\mathbf{M}$ onto $\mathbf{k}_f$ at $\psi = \pm 90$ are therefore unequal.

Information on the spin configuration $\vec{\phi} = (\varphi_1, \varphi_2, \varphi_3)$ is contained in the relative intensity of the peaks and the disproportion of the maxima at $\psi \pm 90°$, which also corresponds to small shifts of the zero-point away from $\psi = 0°$. Fig. S9(c) is a least-squares, i.e. $\chi^2(\varphi_1, \varphi_2, \varphi_3)$, map of fits to the data in which only a global scale factor was adjusted. The $(0, \varphi_2, \varphi_3)$ slice shown here contains all four possible collinear configurations (A:"+ + +",B:"+ − −", C:"+ − +", D:"+ + −"). We find a global minimum at B, a local minimum at A, and strong contradictions with any other configurations. Fig. S9(d,e) further illustrate the $\chi^2$ minima at A and B and their magnetic structures. The corresponding best fits are compared with the data in Panel (b). The structure factor of $Q = (0, 14, 0.5)$ given in Panel (a) explains that this peak becomes very weak, as observed in the data, if the spins at the Eu1 sites are anti-parallel to those at the other sites, i.e. solution B.

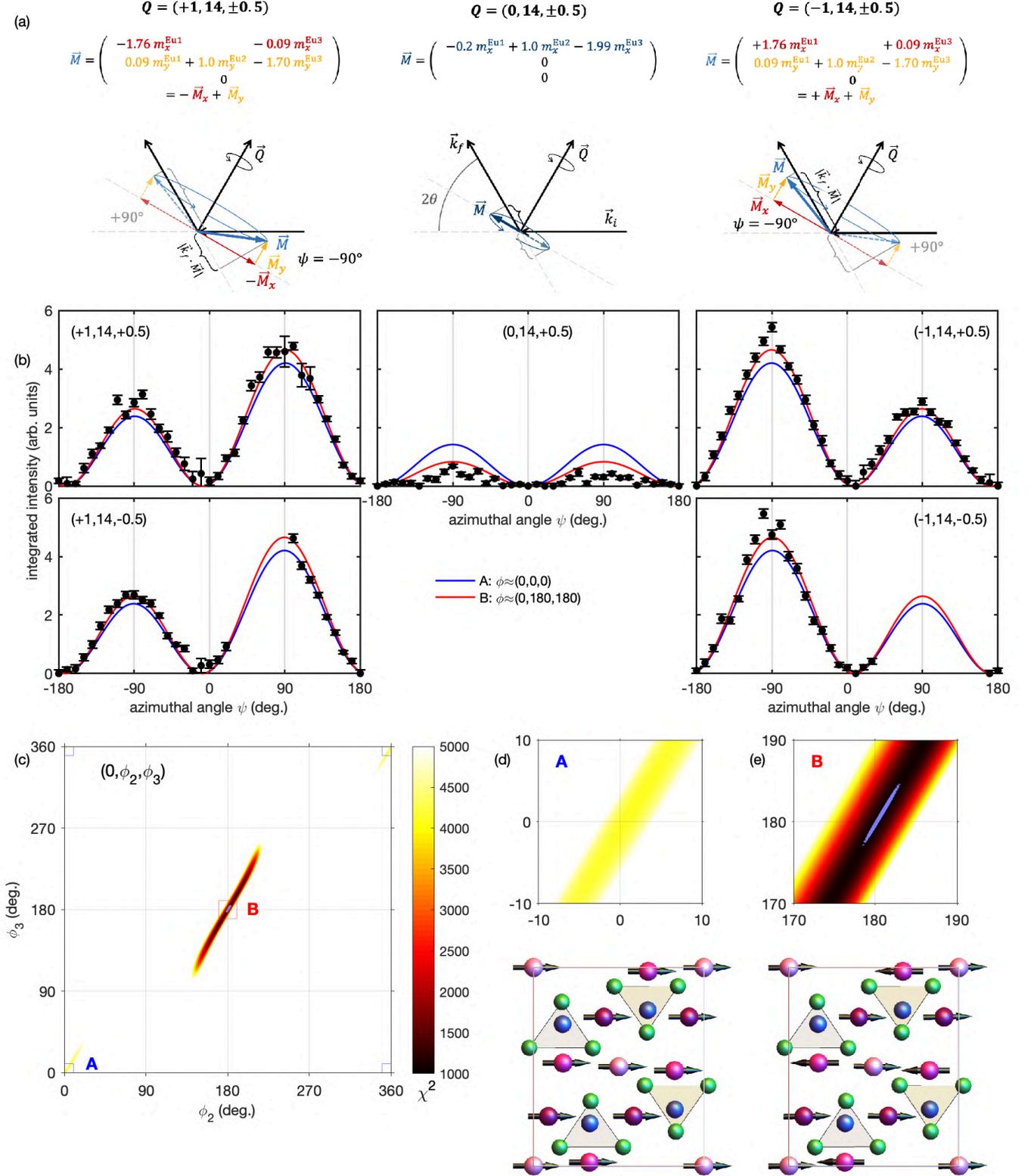

Fig. S9. Analysis of REXS $\sigma\pi'$ azimuthal scans of magnetic peaks close to the specular reflection $Q = (0, 1, 0)$, using $Q = (0, 0, 1)$ as the azimuthal reference ($\psi = 0$). (a) Schematic illustrations of the scattering geometry at $\mathbf{Q} = (h, 14, \pm 0.5)$ magnetic peaks (with $h = -1, 0, 1$). The measured intensity is proportional to $|\mathbf{k}_f \cdot \mathbf{M}|^2$, as indicated by curly brackets. (b) Azimuthal variation integrated intensities for corresponding magnetic Bragg peaks. Lines indicate the characteristics of the "A" and "B" structures. (c) Least-squares ($\chi^2$) map of global fits of the data for any spin configuration $\phi = (0, \varphi_1, \varphi_2)$. Local and global minima are observed for the configurations "A" and "B", respectively. (d,e) Detailed views of the $\chi^2$ maps around these minima with illustrations of their spin configurations. The light blue contour indicates an interval of one standard deviation from the global minimum.

# 8. X-ray absorption spectroscopy

X-ray absorption spectra obtained along REXS measurements confirm that all Eu ions in $Eu_5In_2Sb_6$ are divalent, see Fig. S10

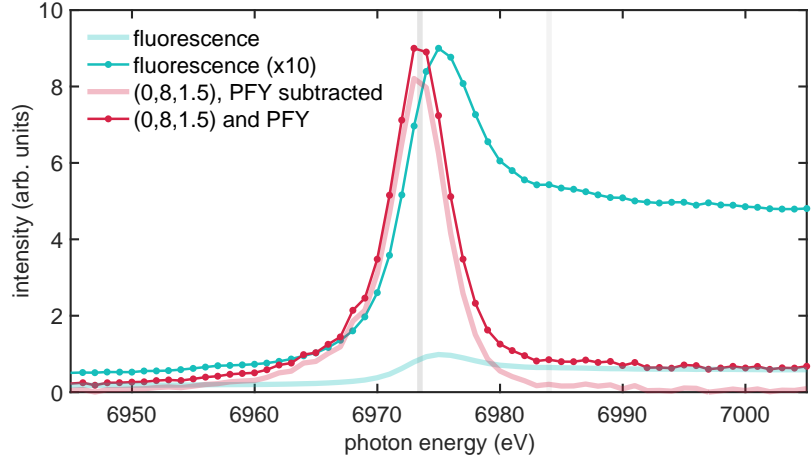

Fig. S10 View of the Eu $L_3$ ($2p_{3/2} \leftrightarrow 5d$) absorption edge as observed at P09 (DESY). The markers show the data of the partial fluorescence yield X-ray absorption spectrum (PFY-XAS) and the intensity of the (8,0,1.5) Bragg peak recorded without polarization analyser (i.e. with fluorescence background intensity). The shaded lines indicate the subtraction of the fluorescence background from the Bragg peak. The resonant energy (6973.5 eV) is marked by a vertical line. The observations rule out a contamination by nonmagnetic $Eu^{3+}$ ($S = 3$, $L = 3$, $J = 0$) ions, for which the XAS edge is shifted to 6984 eV (also marked by a vertical line).

## 9. Muon-spin relaxation

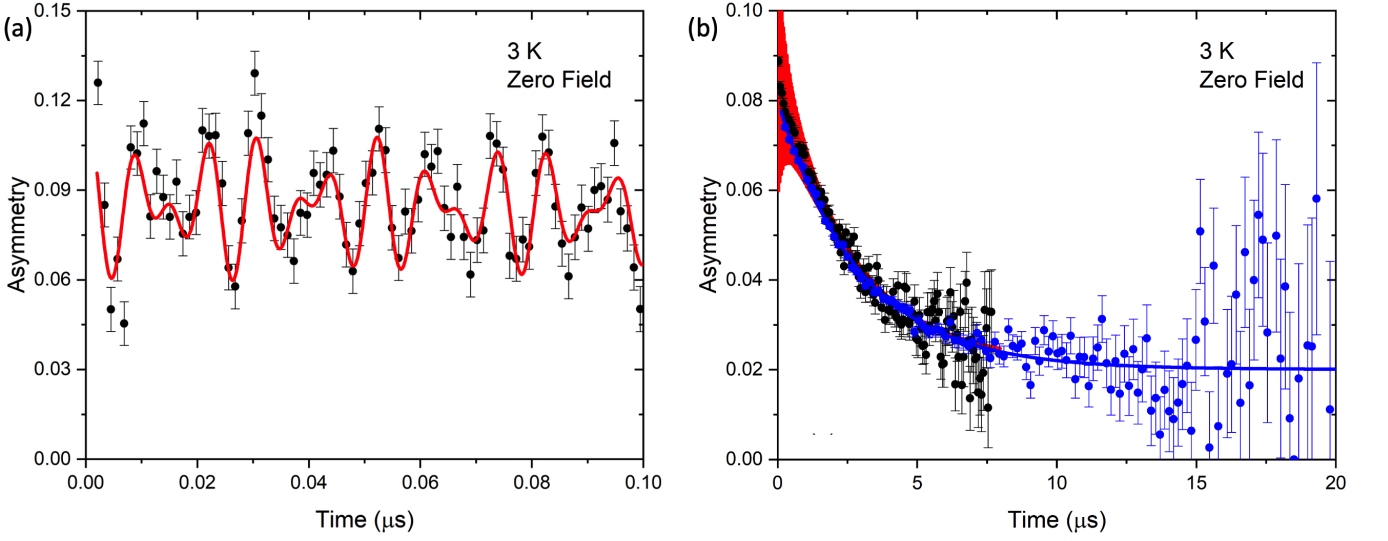

Fig. S11 $\mu$SR data measured in zero applied field at $T = 3$ K (solid points) along with the fit to Eq. 5 (lines), measured using EMU (blue) and GPS (black). (a) Data measured at early times; (b) spectrum plotted out to 20 $\mu$s with greater binning of the data. Note that, on this scale, the GPS fit-line appears as a red-shaded range around the rebinned black markers.

Fig. S11 shows an example of a $\mu$SR spectrum measured in zero applied magnetic field at $T = 3$ K. These data display a combination of complex rapid oscillations and slow relaxation, suggesting the presence of both static magnetism and dynamics on the muon timescale (MHz). We find that the data can be best described by a combination of four components in the asymmetry,

$$
\begin{aligned}
A = \quad & A_{\text{inter}}^{\text{stat}} \quad \mathrm{e}^{-\lambda_{\text{inter}}t} \quad \cos(\phi_{\text{inter}} + \gamma_\mu \mu_0 H_{\text{inter}} t) \quad + \quad A_{\text{inter}}^{\text{dyn}} \quad \mathrm{e}^{-\lambda_{\text{inter}}^{\text{dyn}} t} \\
+ \; & A_{\text{intra}}^{\text{stat}} \quad \mathrm{e}^{-\lambda_{\text{intra}}t} \quad \cos(\phi_{\text{intra}} + \gamma_\mu \mu_0 H_{\text{intra}} t) \quad + \quad A_{\text{intra}}^{\text{dyn}} \quad \mathrm{e}^{-\lambda_{\text{intra}}^{\text{dyn}} t} \quad ,
\end{aligned}
\tag{5}
$$

where $A_i^j$ are the amplitudes of the different asymmetry components. The amplitudes of the two oscillating components, $A_{\text{inter}}^{\text{stat}}$ and $A_{\text{intra}}^{\text{stat}}$, are fixed to be the same. $\lambda_i^j$ are the relaxation rates, $\phi_i$ are phase offsets, $\gamma_\mu = 2\pi \times 135.5$ MHz/T is the muon gyromagnetic ratio, and $\mu_0 H_{\text{inter}}$ and $\mu_0 H_{\text{intra}}$ are the static internal fields, measured in Tesla. The two oscillating components are not resolved in the data measured on EMU. This equation fitted to the data to obtain the parameters displayed in Fig. 3 of the manuscript.

In addition to the zero-field $\mu$SR measurements described in the main text, we also performed longitudinal-field (LF) $\mu$SR measurements on the EMU spectrometer at ISIS, where the magnetic field $H_L$ is applied along the initial direction of the muon spin. These measurements are sensitive primarily to the dynamic character of the magnetism, where changes to the observed muon relaxation rate with applied field can reveal information about the nature of magnetic fluctuations in the material being studied. We performed LF-$\mu$SR measurements at both 25 K and 150 K to check for a possible continued evolution of the magnetism above $T_N$, and found that at both temperatures the LF-$\mu$SR spectra can be well described by two exponentially relaxing components, $A = A_{inter}^{dyn} \exp\left(-\lambda_{inter}^{dyn} t\right) + A_{intra}^{dyn} \exp\left(-\lambda_{intra}^{dyn} t\right)$. Figure S12 shows the field-dependence of these two relaxation rates at 25 K and 150 K.

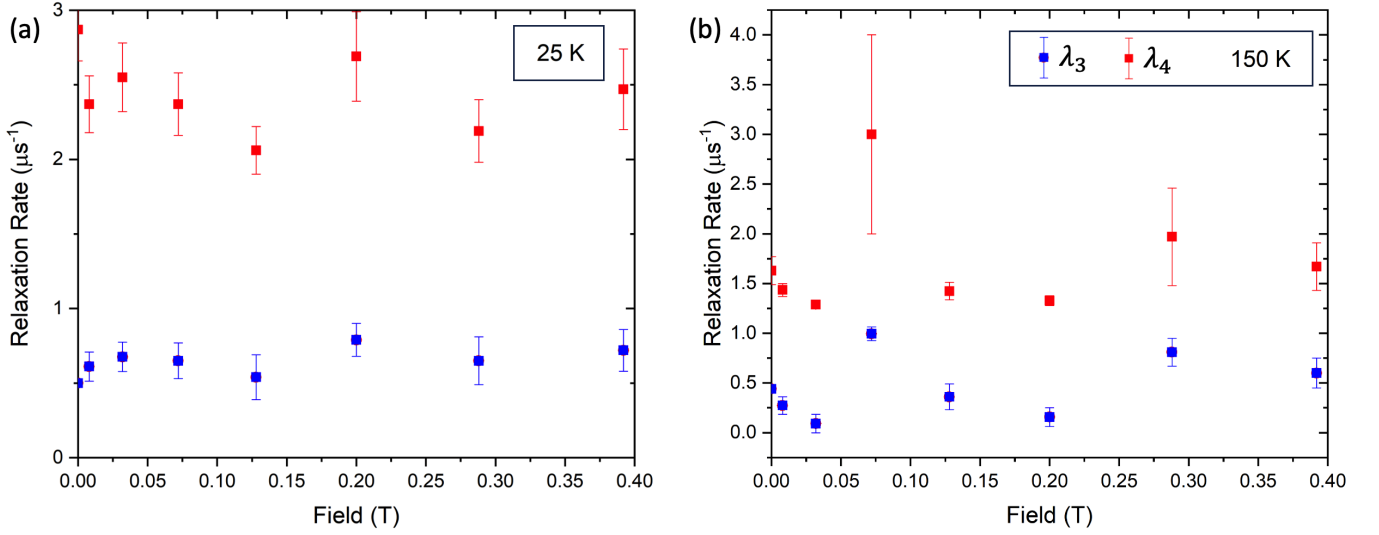

Fig. S12 Dynamic relaxation rates $\lambda_{inter}^{dyn}$ (blue) and $\lambda_{intra}^{dyn}$ (red) for the longitudinal field $\mu$SR data measured on EMU. (a) $T = 25$ K, (b) 150 K.

We do not see a strong trend in the field dependence in the relaxation rate up to 0.4 T at either temperature. This, combined with the observed slow decrease in the relaxation rates with increasing temperature above 20 K seen in Fig. 4(a) of the manuscript, is what would be expected from a system in the paramagnetic state in the fast fluctuation regime. This involves rapid magnetic fluctuations at frequencies greater than the muon Larmor frequency $\gamma_\mu H_L$ at both temperatures and fluctuation rates that continue to increase with increasing temperature, resulting in further decreases to the relaxation rate (due to motional narrowing).

# 10. Nearest-neighbor exchange paths

To consider the effects of magnetic exchange frustration, it may be instructive to consider the nearest-neighbor paths of itinerant exchange interactions between the localized Eu. The distribution of the ten Eu (per unit cell) in the $a$-$b$ plane of $Eu_5In_2Sb_6$ can be interpreted as a network of corner-sharing and corner-and-edge-sharing irregular triangles. This is illustrated in Fig. S12.

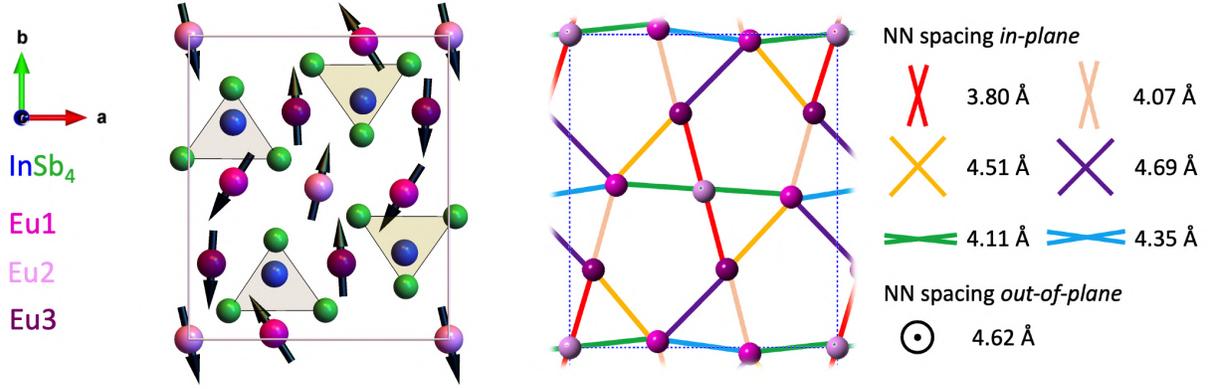

Fig. S13 (left) Crystal structure and magnetic motif of $Eu_5In_2Sb_6$ viewed along the $c$ axis. There are ten Eu ions, all confined to the $a$-$b$ plane. The ions occupy one 2-fold Wyckoff site (Eu2 on $2a$) and two 4-fold Wyckoff sites (Eu1 and Eu3, both on $4g$). This yields an irregular arrangement of Eu ions in the channels surrounding the chains (along $c$) of $InSb_4$ tetrahedra. (right) This arrangement corresponds to 6+1 (in-plane + out-of-plane) distinct nearest-neighbor exchange paths. The distances (at room temperature) range between 3.80 Å and 4.69 Å. Notably, the out-of-plane spacing (i.e. the lattice parameter $c$), 4.62 Å, is intermediate to these values. The Eu2 and Eu3 sites have four "near neighbours" while Eu1 sites have five.

[1] J. Rodríguez-Carvajal, Recent advances in magnetic structure determination by neutron powder diffraction, Physica B: Condensed Matter **192**, 55 (1993).

[2] P. F. S. Rosa, Y. Xu, M. Rahn, J. Souza, S. Kushwaha, L. Veiga, A. Bombardi, S. Thomas, M. Janoschek, E. D. Bauer, M. Chan, Z. Wang, J. Thompson, N. Harrison, P. Pagliuso, A. Bernevig, and F. Ronning, Colossal magnetoresistance in a nonsymmorphic antiferromagnetic insulator, npj Quantum Materials **5**, 52 (2020).



\end{document}